\documentclass{aa}

\usepackage{natbib,twoopt}

\usepackage{hyperref}
\hypersetup{colorlinks=true,allcolors=blue}
\usepackage{txfonts}
\usepackage{graphicx}
\usepackage[export]{adjustbox}
\usepackage{nicefrac}

\usepackage[normalem]{ulem}

\begin{document}

\title{Discovery of a supercluster in the eROSITA Final Equatorial Depth Survey: X-ray properties, radio halo, and double relics }
\titlerunning{eFEDS supercluster}
\mail{vittorio.ghirardini@cfa.harvard.edu}

\author{V. Ghirardini\inst{1}\thanks{e-mail: \href{mailto:vittorio@mpe.mpg.de}{\tt vittorio@mpe.mpg.de}} 
\and E. Bulbul\inst{1} 
\and D. N. Hoang\inst{2}
\and M. Klein\inst{3,16,1}
\and N. Okabe\inst{5,6,7}  
\and V. Biffi\inst{3,4} 
\and M. Br{\"{u}}ggen\inst{2}  
\and M. E. Ramos-Ceja\inst{1} 
\and J. Comparat\inst{1} 
\and M. Oguri\inst{8,9,10} 
\and T. W. Shimwell\inst{14,13}
\and K. Basu\inst{17}
\and A. Bonafede\inst{11,12,2}
\and A. Botteon\inst{13}
\and G. Brunetti\inst{12}
\and R. Cassano\inst{12}
\and F. de Gasperin\inst{2}
\and K. Dennerl\inst{1}
\and E. Gatuzz\inst{1}
\and F. Gastaldello\inst{17}
\and H. Intema\inst{13,15}
\and A. Merloni\inst{1}
\and K. Nandra\inst{1}
\and F. Pacaud\inst{17}
\and P. Predehl\inst{1}
\and T. H. Reiprich\inst{17}
\and J. Robrade\inst{2}
\and H. R{\"{o}}ttgering\inst{13}
\and J. Sanders\inst{1}
\and R. J. van Weeren\inst{13}
\and W. L. Williams\inst{13}
}
\authorrunning{V. Ghirardini et al.}

\institute{
Max-Planck-Institut f{\"{u}}r extraterrestrische Physik, Giessenbachstra{\ss}e 1, D-85748 Garching, Germany \and 
Hamburger Sternwarte, University of Hamburg, Gojenbergsweg 112, 21029 Hamburg, Germany 
\and
Universitaets-Sternwarte Muenchen, Fakultaet fuer Physik, LMU Munich, Scheinerstr. 1, 81679 Munich, Germany 
\and
IFPU - Institute for Fundamental Physics of the Universe, Via Beirut 2, 34014 Trieste, Italy 
\and 
Physics Program, Graduate School of Advanced Science and Engineering, Hiroshima University, 1-3-1 Kagamiyama, Higashi-Hiroshima, Hiroshima 739-8526, Japan
\and
Hiroshima Astrophysical Science Center, Hiroshima University, 1-3-1 Kagamiyama, Higashi-Hiroshima, Hiroshima 739-8526, Japan
\and
Core Research for Energetic Universe, Hiroshima University, 1-3-1, Kagamiyama, Higashi-Hiroshima, Hiroshima 739-8526, Japan 
\and
Research Center for the Early Universe, The University of Tokyo, 7-3-1 Hongo, Bunkyo-ku, Tokyo 113-0033, Japan 
\and
Department of Physics, The University of Tokyo, 7-3-1 Hongo, Bunkyo-ku, Tokyo 113-0033, Japan 
\and
Kavli Institute for the Physics and Mathematics of the Universe (Kavli IPMU, WPI), The University of Tokyo, 5-1-5 Kashiwanoha, Kashiwa, Chiba 277-8582, Japan 
\and Dipartimento di Fisica e Astronomia, Università di Bologna, via P. Gobetti 93/2, I-40129 Bologna, Italy
\and INAF - IRA, via P. Gobetti 101, I-40129 Bologna, Italy
\and Leiden Observatory, Leiden University, PO Box 9513, 2300 RA Leiden, The Netherlands
\and ASTRON, the Netherlands Institute for Radio Astronomy, Postbus 2, NL-7990 AA Dwingeloo, The Netherlands
\and 
International Centre for Radio Astronomy Research - Curtin University, GPO Box U1987, Perth, WA 6845, Australia
\and 
Faculty of Physics, Ludwig-Maximilians-Universit{\"a}t,
Scheinerstr. 1, 81679, Munich, Germany
\and 
Argelander-Institut f{\"{u}}r Astronomie (AIfA), Universit{\"{a}}t Bonn, Auf dem H{\"{u}}gel 71, 53121 Bonn, Germany
\and
INAF - IASF Milano, via Bassini 15, I-20133 Milano, Italy
}

\mail{vittorio@mpe.mpg.de}
\abstract
{}
{
We examine the X-ray, optical, and radio properties of the member clusters of a new supercluster discovered during the SRG/eROSITA Performance Verification phase.
} 
{
We analyzed the 140~deg$^2$ eROSITA Final Equatorial Depth Survey (eFEDS) field  observed  during the Performance Verification phase to a nominal depth of about 2.3~ks. In this field, we detect a previously unknown supercluster consisting of a chain of eight galaxy clusters at $z \sim$ 0.36. The redshifts of these members were determined through Hyper Suprime-Cam photometric measurements.
We examined the X-ray morphological and dynamical properties, gas, and total mass out to $R_{500}$ of the members and compare these with the same properties of the general population of clusters detected in the eFEDS field. 
We further investigated the gas in the bridge region between the cluster members for a potential WHIM detection. We also used radio follow-up observations with LOFAR and uGMRT to search for diffuse emission and constrain the dynamic state of the system.}
{We do not find significant differences between the morphological parameters and properties of the intra-cluster medium  of the clusters embedded in this large-scale filament and those of the eFEDS clusters. We also provide upper limits on the electron number density and mass of the warm-hot intergalactic medium as provided by the eROSITA data. These limits are consistent with previously reported values for the detections in the vicinity of clusters of galaxies. In LOFAR and uGMRT follow-up observations of the northern part of this supercluster, we find two new radio relics and a radio halo that are the result of major merger activity in the system. }
{These early results show the potential of eROSITA to probe large-scale structures such as superclusters and the properties of their members. 
Our forecasts show that we will be able to detect about 450 superclusters, with approximately 3000 member clusters located in the eROSITA\_DE region at the final eROSITA all-sky survey depth, enabling statistical studies of the properties of superclusters and their constituents embedded in the cosmic web.
 }


\maketitle
\section{Introduction}
Cosmic structures evolve hierarchically from high-density peaks in the primordial density field, and form galaxies, galaxy groups, and clusters of galaxies under the action of gravity. In the complex large-scale structure formation scenario, these galaxies, groups, and clusters are connected to each other via filamentary structures called the cosmic web \citep{springel+05}, and form large superclusters \citep{bond+1996,einasto+1997}. Due to the large crossing times, superclusters are neither virialized nor relaxed, although individual structures located within superclusters, such as galaxy clusters, can be gravitationally bound and virialized \citep{Pearson+13}. Superclusters contain a variety of structures with a range of masses, from massive and dense clusters of galaxies to low-density bridges, filaments, and sheets of matter, and are therefore ideal laboratories in which to study the physical processes that affect the evolution of member galaxies, groups, and clusters \citep[e.g.,][]{springel+05}. It has been suggested that the supercluster environment strongly influences the evolution of the properties of its constituents \citep[e.g.,][]{aragon-calvo+2010, einasto+2011}. For instance, simulations predict that the shape of clusters in superclusters is predominantly elongated, and the elongation is typically along the filament direction \citep[e.g.,][]{Basilakos+06,Lee+07}.  Additionally, comprising
a significant amount of baryons in the form of galaxies and diffuse gas, the filaments connecting clusters of galaxies have typical diameters of approximately a few megaparsec, and coherence lengths of the order of $\sim$~5~Mpc but can extend up to $\sim$~20-25~Mpc \citep[e.g.,][]{Dolag+05}. Detections of the baryons located in the warm-hot intergalactic medium (WHIM) locked in the filaments could reveal important clues regarding the missing baryon problem in the low-redshift Universe \citep{Shull+2012,Nicastro+18}.

A number of superclusters have been found in deep optical surveys of galaxies \citep[e.g., 2dF Galaxy Redshift Survey, 2MASS Redshift Survey, and SDSS-DR13;][]{ colless+2003, huchra+2012, alparslan+2014, Santiago-Bautista+2020}. In X-rays, the first flux-limited supercluster catalog was compiled by \cite{Chon+13}. Based on the \texttt{REFLEX II} cluster sample \citep{Chon+12}, they located 164 superclusters, of which only a couple are above a redshift of 0.35. 
Although a number of multiwavelength supercluster catalogs exist in the literature, only a few superclusters and their members have been observationally studied in depth, in particular the Shapley supercluster at low redshift, and two high-redshift clusters.

The Shapley supercluster, discovered by \cite{Shapley+30}, is one of the most studied superclusters of galaxies in the sky. It is a concentration of more than 20 galaxy clusters in a volume of about $10^{-3}\ \textrm{Gpc}^3$, and about 20~deg$^2$ in the plane of the sky at a redshift of 0.039.  \cite{Ettori+00} combined \texttt{ROSAT PSPC} and \textit{BeppoSAX} X-ray data to study three members of the Shapley supercluster  in
detail using resolved spatial and spectral analysis. These latter authors concluded that A3562, a member of the Shapley supercluster, is not relaxed and shows evidence of merging activity. At high redshifts ($z>0.4$), the number of known superclusters is extremely small, and even fewer have been investigated in detail.
\cite{Horner+03} studied the supercluster MS0302 at z = 0.42, which is composed of
three massive galaxy clusters. This system was discovered in an X-ray follow up observation using the Einstein X-ray observatory of optically detected cluster members. 
\cite{Horner+03} measured the X-ray properties, temperature, and luminosity of the members of the system. 
Recently, \cite{Adami18} presented the discovery of 35 superclusters found in the 50 deg$^2$ XXL survey \citep{XXL}.
Of these clusters, \cite{XXL_supercluster} performed a detailed study of a supercluster at z = 0.43 consisting of six galaxy clusters.
These latter authors determined the temperature, luminosity, and total mass of its members using their internally calibrated weak lensing mass--X-ray temperature scaling relation. From their morphological analysis, they find that two of these six clusters appear to be disturbed, indicating that they are likely in a merging state. {\it Suzaku} observations of A1689 and A1835 \citep{2010ApJ...714..423K,2013ApJ...766...90I} were used to investigate correlations between the X-ray properties in cluster outskirts ($R_{500}<R<R_{200}$) and their surrounding large-scale structure, revealing that the  X-ray temperatures in the outskirts of the regions connected to the filamentary structure are higher than those connected to void regions.

Detailed examination of the connecting bridges of the member clusters of galaxies in superclusters has been an active area of research because of its connection with the missing baryon problem. Detections of the warm-hot intergalactic medium in these regions is quite challenging with current X-ray instrumentation because of the relatively low density ($<10^{-4}$ particle cm$^{-3}$, corresponding to an over-density of 10-100 times the cosmic value\footnote{The critical density of the Universe: $\rho_c = \frac{3 H^2(z)}{8 \pi G}$}) and low temperature ($10^{5}-10^{7}$~K) of this gas. 
Recent dedicated deep X-ray observations suggest the presence of such gas in the interconnecting bridges or filaments between clusters of galaxies, such as for example A3391/95 \citep{Tittley+01,Sugawara+17,Alvarez+18}, A222/3 \citep{Werner+08}, A2744 \citep[][]{eckert+15}, A1750 \citep[][]{Bulbul+16}, and A133 \citep[][]{connor+18,connor+19}. An alternative way to detect this low-density gas is through the thermal Sunyaev Zel'dovich (tSZ) effect. \citet{planck+2013} reported their detection of such a gas in the A399-A401 cluster pair with an estimate for the gas temperature of $kT =8\times10^{7}$~K and for the electron density of $n_{e} =3.7\times 10^{-4}$ cm$^{-3}$.  
It should be noted that these potential detections probe the densest and hottest ends of the WHIM, where the intracluster gas interacts with the colder primordial low-density WHIM gas. Recently, by stacking the Planck Compton $y$-parameter map of the tSZ signal of galaxy pairs, \citet{degraaff+2019} reported a $2.9\sigma$ detection of the WHIM gas with a gas density of $\rho= 5.5\pm2.9\ \overline{\rho_{b}}$, where $\overline{\rho_{b}}$ is the mean matter density of the Universe, and a gas temperature of $kT =(2.7\pm 1.7$)\ $\times10^{6}$~K. Recently, diffuse synchrotron radio emission was detected in the region connecting the pairs A1758N/A1758S \citep{Botteon+18,Botteon2020a} and A399/A401 \citep{Govoni+2019} using the LOw Frequency ARray (LOFAR; \citealt{Haarlem2013}). The origin of the radio synchrotron emission in radio bridges is not well understood, but turbulence generated by the stochastic acceleration of relativistic electrons could be an explanation for the observed diffuse radio emission in A399--A401 \citep{Brunetti+2020}.

Statistical multi-wavelength studies of the properties of member structures embedded in superclusters are crucial to the development of our understanding of the evolution of the large-scale structure. Here we report the discovery of a new supercluster at a redshift of 0.36 in the eROSITA Final Equatorial Depth Survey (eFEDS) performed during the Performance Verification (PV) program. 
{ eFEDS is a 140 deg$^2$ field located in an equatorial region, with R.A. from $\sim$127 to $\sim$145, and Dec. from $\sim$-2 to $\sim$5. It was observed in scanning mode by eROSITA with nominal exposure of about 2.3 ks.}
In this paper, we examine the X-ray and radio properties of the member clusters of galaxies in their large-scale environment and the WHIM gas in the interconnecting region joining the observations from eROSITA in the X-ray, Hyper Suprime-Cam (HSC) in the optical, and LOFAR and uGMRT in the radio band.
We structure the paper as follows: in Sect.~\ref{sec2} we describe the new eROSITA X-ray observations and analysis, and HSC optical data. We provide our results on the X-ray  properties of the member clusters and WHIM gas in Sect~\ref{sec3}. We describe the radio results in Sect.~\ref{sec:radio}. Our conclusions and a summary are given in Sect~\ref{sec:sec4}. Throughout this paper we assume a concordance $\Lambda$CDM cosmology with $\Omega_m = 0.3$, $\Omega_\Lambda = 0.7$, and $H_{0} = 70 \textrm{ km s}^{-1}\ \textrm{Mpc}^{-1} $.

\section{Multi-wavelength data, their analysis, and reduction}
\label{sec2}
\subsection{eROSITA X-ray data reduction}
The extended ROentgen Survey with an Imaging Telescope Array (eROSITA,Predehl et al. 2020 in press) onboard the Spectrum-Roentgen-Gamma (SRG) mission (Sunyaev et al. 2020) was launched on July 13$^{\rm }$, 2019. The large effective area (1365 cm$^2$ at 1 keV), large field of view (FoV, 1~deg diameter), good spatial resolution (half energy width of 28~arcsec averaged over the FoV at 1.49~keV) and spectral resolution ($\sim 80$~eV full width half maximum at 1~keV) of eROSITA allow unique survey science capabilities by scanning large areas of the X-ray sky 
quickly and efficiently \citep{merloni+12}. 

 The 140~deg$^2$ eFEDS field was uniformly scanned during the Performance Verification phase resulting in a nominal exposure of about 2.3 ks (unvignetted) over the field, which is similar in depth to the final exposure that will be reached in 4 years in equatorial fields in the eROSITA All-sky survey. eFEDS, currently the largest contiguous X-ray survey at this depth, has been designed to demonstrate the unique survey science capabilities of eROSITA. eFEDS will enable calibration of key mass scaling relations, combining X-ray properties with 
 weak lensing masses of detected groups and clusters
 obtained from the HSC on the Subaru telescope. 
 eFEDS observations will enable eROSITA to test  the  projections for cosmological parameters in the all-sky survey.

The eFEDS data were acquired by eROSITA over 4 days, between November 4$^{\rm }$ and 7$^{\rm }$ 2019. These data were processed by the eROSITA Standard Analysis Software System (eSASS, Brunner et al. in prep.). 
The eSASS pipeline produces calibrated event lists for each eROSITA telescope module (TM) by applying pattern recognition and energy calibration, accompanied by the determination of good time intervals, dead times, corrupted events and frames, and bad pixels.
Using star-tracker and gyro data, celestial coordinates are assigned to each reconstructed X-ray photon. After this step, photons can be projected onto the sky so that  images and exposure maps can be produced. 
In this work, we select all valid pixel patterns, that is, single, double, triple, and quadruple events, but  only use photons detected at off-axis angles  $<=~$30 arcmin (i.e., we remove photons in the corners of the square CCDs where the vignetting and PSF calibration is currently less accurate).

The source detection procedure is performed on the merged data from all seven eROSITA telescope modules. The detection is based on a sliding-cell method. In a first step, the algorithm scans the X-ray image with a local sliding window, which identifies enhancements above a certain threshold. The detected candidate objects are then excised from the images. The resulting source-free images are used to create background maps via adaptive filtering. The sliding window detection is then repeated, but this time using the created background map to search for signal excess with respect to the background, producing another candidate source list. For each source candidate, a maximum likelihood point spread function (PSF)-fitting algorithm determines the best-fit source parameters and detection and extension likelihoods. Applying this algorithm on the eFEDS data, using images in the $0.2-2.2$ keV energy band, an extent likelihood threshold of 8, and source extension threshold of $60$~arcsec, we detect $450$~extended sources in total (full details of the eFEDS data processing, analysis, and source detection will be presented in future work). 

\begin{figure*}
\includegraphics[width=0.5\textwidth]{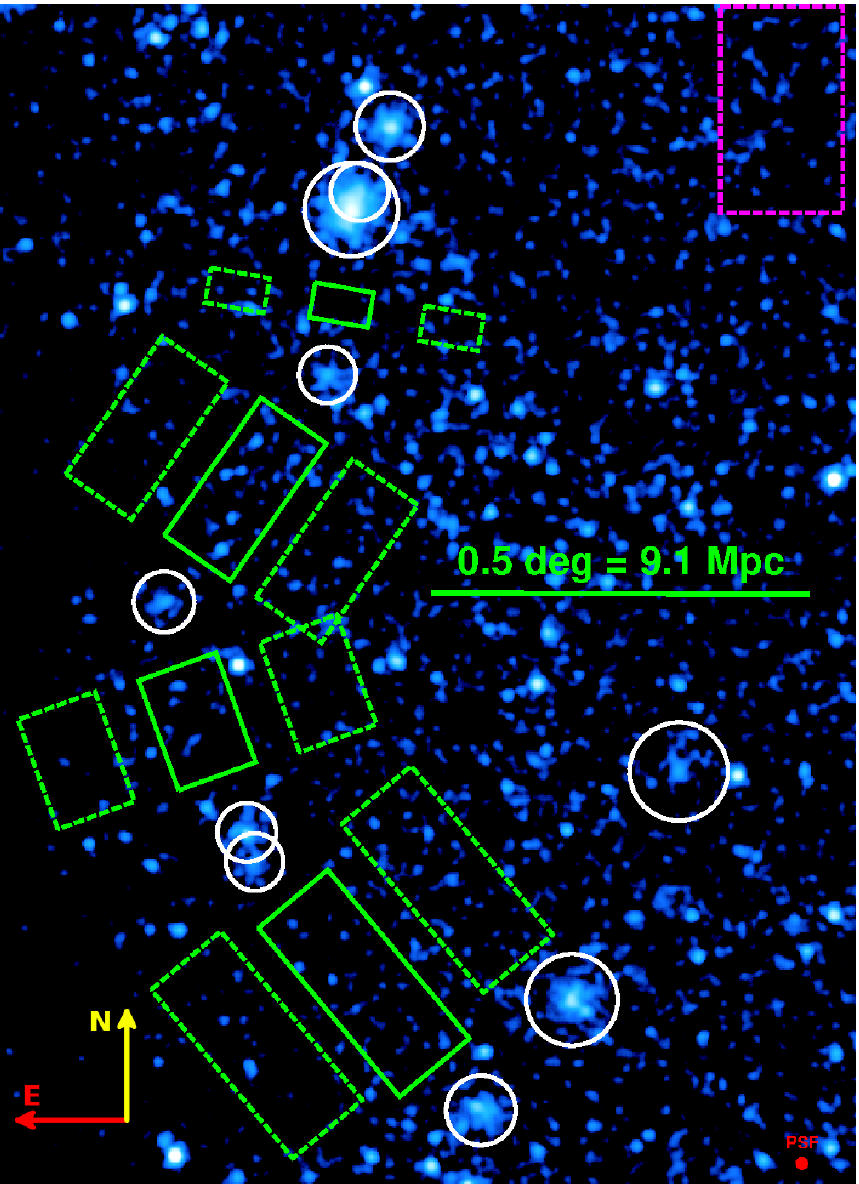}~
\includegraphics[width=0.5\textwidth]{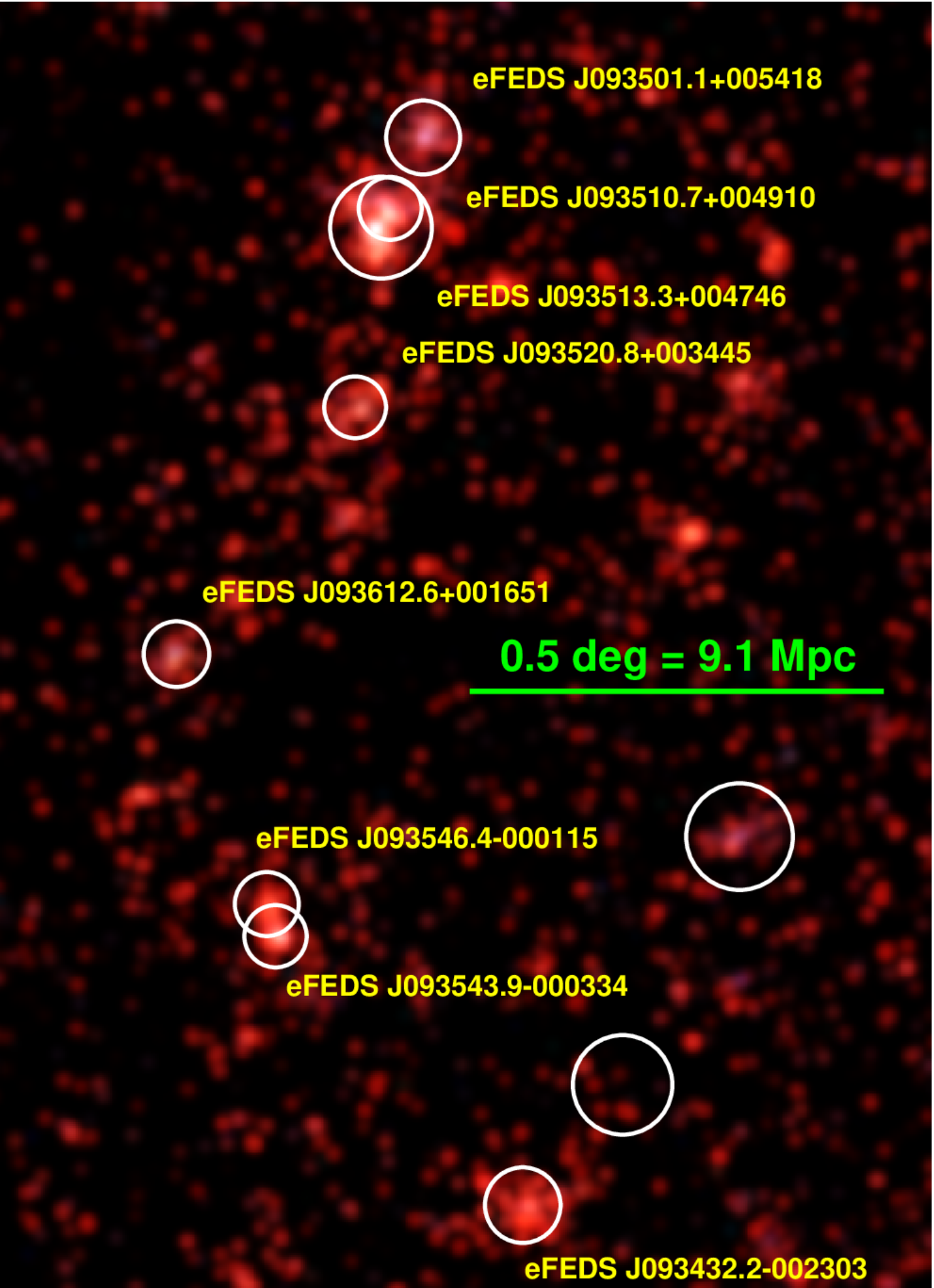}~
\caption{\emph{Left}: eROSITA X-ray image in the soft [0.5-2.0] keV energy band, smoothed with a Gaussian of $\sigma$=8 arcsec.
The solid green boxes represent the location of the region within which we look for emission from the filament connecting these clusters. Dashed green boxes represent the local background for spectral analysis. In order to minimize bias toward cosmic variance we repeat the analysis using the magenta box as an alternative background region. 
\emph{Right}: Color image of the galaxy density map at redshift of 0.36 from HSC.
In both panels, we mark the location of the clusters detected by eROSITA  with white circles, their size representing the estimated $R_{500}$.
}
\label{fig1}
\end{figure*}
\subsubsection{eROSITA Imaging analysis}
\label{sec:spatial}

We extract images and exposure maps in the 0.5--2.0~keV energy band using eSASS tasks \texttt{evtool} and \texttt{expmap,} respectively. The resulting soft band image of the zoomed 1.1~$\times$~1.5~deg southwest region of the eFEDS field, where the supercluster is located, is displayed in the left panel of Figure~\ref{fig1}; see Sect.~\ref{sec:supercluster_definition} for details on how the supercluster was found.
We note that the cluster member of the supercluster has not been discovered by any previous X-ray observation from \emph{Chandra} or \emph{XMM-Newton}.
We also direct the reader to Figure~\ref{fig_sens_exp} for the eROSITA sensitivity and exposure map of the area. 

Faint point sources (with less than 0.1 count per second) detected in the images are excised from further analysis. The masking radius of the faint point sources in the FoV is estimated using the observed point source profile. This profile is obtained multiplying the  PSF profile by the point source count rate, and the radius where the point source profile is consistent with the background level within the $1\sigma$-level is marked as the excision region. Bright point sources are, on the other hand, added as extra model components to the total model 
(see below) as delta functions convolved with the PSF to fully account for the wings of the PSF.
 
To derive the cluster properties,
we start by constructing a likelihood function that is used to fit the image obtained from the eROSITA  observations. We project a cluster model \citep[][]{vikhlini+06} in the 2D image plane. We then convolve this model image with the PSF of eROSITA, which is calibrated by the eSASS team and available in the eSASS calibration database (CALDB). The instrumental background (particle induced background and camera noise) model 
folded with the unvignetted exposure map, and the sky background (including the cosmic X-ray background and the soft background component from the Galactic halo and local bubble) folded with the vignetted exposure map are added to the total model. 
We then fit the image obtained from the eROSITA observations with the  model image in 2D using the Monte Carlo Markov Chain (MCMC) code \textit{emcee} \citep{emcee} to find the best-fit model parameters \citep{vikhlini+06}. We assume a Poisson log-likelihood function $N_i - \mu_i \log N_i$, where $N_i$ are the model predicted counts, and $\mu_i$ are the observed counts in each pixel of the image. We note that multiple clusters located in the field are modeled simultaneously.

 The gas mass profile is calculated by integrating the gas density\footnote{The fitting of the images is a true density once we consider the emissivity of the gas, which is determined, as in the case of luminosity, by making use of the spectral information.} over the cluster volume. To measure the luminosity we integrate the surface brightness along the line of sight and convert it to luminosity using the temperature obtained from spectral analysis (see Section~\ref{sec:spectral}). The uncertainties are recovered by bootstrapping through 1000 iterations, parsing a set of surface brightness and temperature values in MCMC chain outputs.

\subsubsection{eROSITA X-ray spectral analysis}
\label{sec:spectral}
 \begin{figure}[t]
     \centering
     \includegraphics[width=0.5\textwidth]{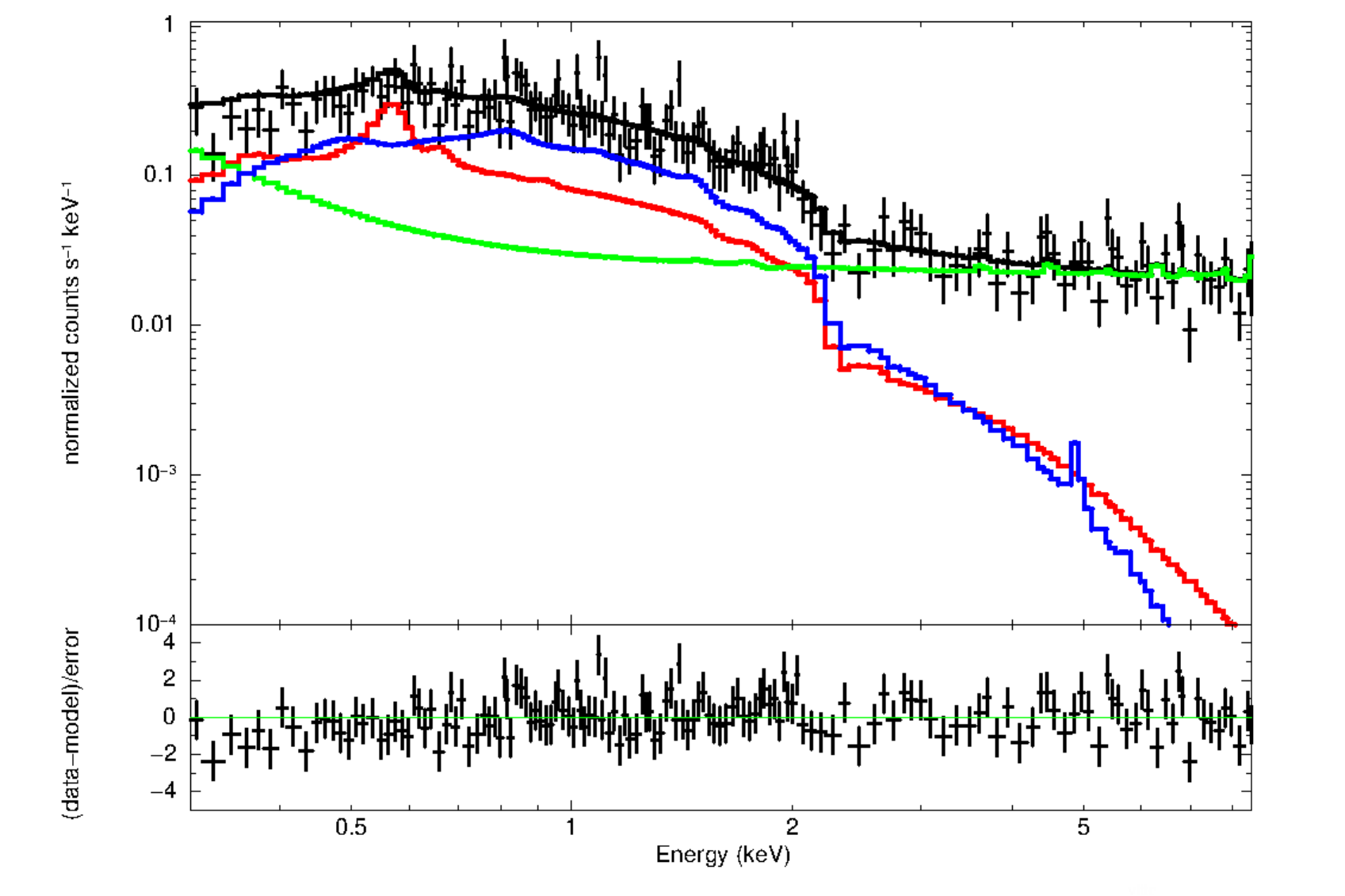}
     \caption{Example of an eROSITA spectrum from all TMs combined{ (for eFEDS~J093510.7+004910)}. Black data points represent the spectral data, and the black line, line, line, and red lines represent the total model, the cluster model, the instrumental background, and the X-ray background,  respectively.}
     \label{fig:spectrum}
 \end{figure}

The spectra, response matrices (RMFs) and effective areas (ARFs) for the source and background regions are extracted using the eSASS task \texttt{srctool} with the current version of the eROSITA CALDB. Taking advantage of the large FoV of eROSITA to model the local Galactic background and the cosmic X-ray background due to unresolved point sources, we extract a background spectrum from an annulus region between 3 and 4$\ R_{500}$\footnote{$R_{500}$ is obtained iteratively, starting from a value, computing the luminosity, and estimating a new $R_{500}$ using the scaling relation. This process is repeated several times until convergence is reached.} surrounding the centroid of each cluster.

Throughout the spectral analysis we adopt the \cite{lodd} abundance table and \textit{cstat} likelihood, the implementation in \texttt{xspec} of Poisson statistics. An example spectrum is shown in Figure~\ref{fig:spectrum}. The fitting band is restricted to the 0.8--9~keV band for the TMs affected by the light leak (TM5 and TM7; see Predehl et al. 2020), while the full 0.3-9~keV band is used for all other TMs.
The cluster temperature, metallicity, and normalization are determined by fitting the spectra with an absorbed thermal \texttt{apec} model with ATOMDB version~3.0.9 \citep{foster+2012} within the software package \texttt{xspec 12.10.1} \citep{xspec}. The absorption model TBabs is used to account for the Galactic absorption \citep{tbabs}.

Source and background spectra for the individual TMs are fitted simultaneously with their appropriate responses and effective areas. Our total model includes an absorbed thermal model for the cluster itself, two absorbed thermal models for the galactic halo, one unabsorbed thermal model for the local hot bubble, and absorbed power-law model for the cosmic X-ray background due to the unresolved point sources. These models were folded through the appropriate effective area and response. We also added extra power-law models and a few Gaussian emission lines to take into account the instrumental background. This component is not folded through the ARF file as the highly energetic cosmic-ray (CR) particles do not get focused through the telescope mirror \citep{Bulbul+2020}. The shape of the instrumental background has been fixed to the values obtained from the analysis of the filter-wheel-closed data and only the normalizations of the instrumental background components were left free during the fit. To explore a wide parameter space, we use the Goodman-Weare MCMC code implemented within \texttt{xspec} with 300k iterations.

\subsection{Hyper Suprime-Cam optical data}
\label{sec:HSC}

The Subaru telescope is a large 8.2-meter telescope located on Mauna Kea, Hawaii.
The Hyper Suprime-Cam \citep[HSC;][]{miyazaki18a} installed on the Subaru telescope is a wide-field camera with a 1.5~deg diameter FoV. With the large aperture of the primary mirror and its excellent image quality, the HSC is an ideal instrument for wide-field optical imaging surveys. To fully exploit its survey capability, a wide-field survey named the Hyper Suprime-Cam Subaru Strategic Program \citep[HSC-SSP;][]{aihara18a,aihara18b,aihara19a} is being conducted, which  uses a total of 330 nights of observing time at the Subaru telescope. The Wide layer of HSC-SSP plans to cover $\sim 1 400$~deg$^2$ of the sky with five broad bands ($grizy$) with a 5$\sigma$ limiting magnitude of $r_{\rm lim}\sim 26$. The deep images and the wide wavelength coverage of HSC-SSP allow us to find many optically selected clusters and to accurately measure their photometric redshifts out to $z\sim 1.1$ \citep{oguri18}.
The $i$-band images in the Wide layer are taken under good seeing conditions ($\sim 0\farcs5-0\farcs7$) and are used for weak lensing shape measurements \citep{mandelbaum18}. The survey data are reduced using a dedicated software pipeline \citep{bosch18}. HSC-SSP started in 2014 and plans to finish its survey in 2021.

Ninety percent of the eFEDS area is covered by the latest internal data release of HSC-SSP (S19A). From the HSC-SSP optical data, we measure the photometric redshift for all the extended sources detected by the eROSITA telescope in eFEDS using the recently developed code Multi-Component Matched Filter cluster confirmation tool \citep[MCMF][]{Klein+18,Klein+19}, which allows us to confirm X-ray candidates, to measure the cluster photometric redshift accurately with a relative uncertainty of just 0.6\%, and to measure the cluster richnesses.
The version of MCMF used in this work is very similar to that used in previous publications, with the only adaptations being to the HSC photometric
systems. Known clusters with spectroscopic redshifts over the full HSC-SSP footprint were used for calibrations of MCMF. The HSC data confirmed and measured the redshifts for 338 clusters, thus confirming more than 80\% of the extended sources found by the X-ray pipeline. Besides confirmation based on cluster richness and redshifts, MCMF creates weighted galaxy density maps to identify cluster positions. A version of such a map is shown in the right panel of Figure~\ref{fig1} for a redshift of $z=0.36$.

\subsection{Radio observations}
\label{sec:radio}
\subsubsection{LOFAR}
\label{sec:lofar}
The eFEDS field was observed with LOFAR using High Band Antennas (HBA; project: LC13\_029). The observations with the frequency coverage between 120 and 168~MHz were performed using 74 stations (i.e. 48 split core, 13 remote, and 13 international stations). The  total observing time of 100 hours was spent on six pointings that  entirely cover the eFEDS field. Each pointing was observed in chunks of four hours, bookended by ten-minute observations of calibrators (i.e., 3C~196 and 3C~295). Five pointings were observed for a total duration of 16 hours each. The other westernmost pointing that includes the supercluster northern region (i.e., eFEDS~J093513.3+004746 and eFEDS~J093510.7+004910) and  was observed for 20 hours in total.

We adapt the scheme of the LOFAR Two-meter Sky Survey (LoTSS; {\citealt{Shimwell2017,Shimwell2019}}) to calibrate the LOFAR data recorded with the core and remote stations (international stations are not used in this study). The data-reduction procedure includes the correction for the direction-independent and direction-dependent effects that are caused by the instrument and the ionosphere. We make use of the latest version of the pipelines  \texttt{PREFACTOR}\footnote{\url{https://github.com/lofar-astron/prefactor}} \citep{VanWeeren2016a,Williams2016a,deGasperin2019} and \texttt{ddf-pipeline}\footnote{\url{https://github.com/mhardcastle/ddf-pipeline}} {\citep{Tasse2014a,Tasse2014b,Smirnov2015,Tasse2018,Tasse2020}} for the direction-independent and direction-dependent calibration, respectively. The flux scale of the LOFAR data is calibrated according to the \cite{Scaife2012} scale using the 3C~196 model that has a total flux density of 83.1~Jy at 150~MHz. 

To further improve the fidelity of the LOFAR image in the region of eFEDS~J093513.3+004746 and eFEDS~J093510.7+004910, we perform self-calibration loops on the ``extracted'' data that contain the sources within a  25-arcmin square region centered on the cluster system eFEDS~J093513.3+004746 and eFEDS~J093510.7+004910. The region size is required to be large enough to have sufficient flux density for self-calibration convergence. The extraction also includes the correction for the attenuation of the LOFAR station beam. For details of the extraction and self-calibration, we refer to {\cite{VanWeeren2020a}}.  

To obtain final images of the extracted datasets, the LOFAR calibrated data are deconvolved with \texttt{multi-scale} and \texttt{joint-channel} options in \texttt{WSClean}\footnote{\url{https://sourceforge.net/projects/wsclean/}} \citep{offringa2014,offringa2017} to improve the fidelity  of faint extended emission in the wideband data. To enhance extended emission at different scales, the final LOFAR images are obtained with Briggs' weightings (i.e., \texttt{robust$=-0.25$} and 0.25 with compact sources subtracted). The high-resolution image we made with Briggs' weightings (\texttt{robust$=-0.25$}) resolves the source structure to a resolution of $14.7\,{\rm arcsec}\times7\,{\rm arcsec}$. The properties of the images are given in Table \ref{tab:image_para}.  

\subsubsection{uGMRT}
\label{sec:gmrt}

Follow-up observations of the cluster system eFEDS~J093513.3+004746 and eFEDS~J093510.7+004910 were performed with the upgraded GMRT (uGMRT) Band 3 with 30 antennas on May 31, 2020, for a duration of 6.5 hours (Proposal Code: 38\_004). For calibration purposes, radio sources 3C~147 and 3C~286 were observed in the beginning, the middle, and the end of the target observations. 

We split the uGMRT data into chunks of 33.3~MHz each and used only seven high-frequency chunks (i.e., 266.9--500~MHz) where the data are less affected by the ionospheric conditions. Each chunk is separately processed with the Source Peeling and Atmospheric Modeling (\texttt{SPAM}) package \citep{Intema2009a,Intema2017}. The primarily purpose of the pipeline is to correct for the direction-dependent effects causing the ionospheric phase delay in the data. The flux scale of the uGMRT data is set following the scale described in \cite{Scaife2012}. The calibrated datasets were combined and self-calibrated through phase only and amplitude-phase rounds with the Common Astronomy Software Applications (\texttt{CASA}) package\footnote{ \url{https://casa.nrao.edu/}} to improve the image fidelity. Finally, the calibrated data are deconvolved with multiscale-multifrequency (\texttt{MS-MFS}), multiple Taylor terms (\texttt{nterms=2}), and W-projection options to take into account the complex structure of the extended sources, the wide bandwidth of the data, and the noncoplanar effects of the array \citep{Cornwell2005,Cornwell2008b,Rau2011}. Using Briggs' weighting scheme (\texttt{robust$=-0.25$}) we obtained a high-resolution image, that is, $6.9\,{\rm arcsec}\times5.1\,{\rm arcsec}$ (see Table \ref{tab:image_para} for image properties). Final images are corrected for the uGMRT primary beam with a model of an eighth-order polynomial function,\footnote{A \texttt{CASA} script for the beam correction is available on \url{https://github.com/ruta-k/uGMRTprimarybeam}}

\begin{equation}
B(x) = 1+ \left( \frac{a}{10^3}\right)x^2+ \left( \frac{b}{10^7}\right)x^4 + \left( \frac{c}{10^{10}}\right)x^6 + \left( \frac{d}{10^{13}}\right)x^8,
\label{eqn:beam}
\end{equation}

\noindent where $x$ is the separation (in arcmin) from the pointing center multiplied by the observing frequency (in GHz) and the coefficients for the band 3 are  $a=-2.939$, $b=33.312$, $c=-16.659$, and $d=3.066$.

\subsubsection{Radio spectral imaging}
\label{sec:spx}
We combine the LOFAR and uGMRT data to make spectral index maps of the radio emission from eFEDS~J093513.3+004746 and eFEDS~J093510.7+004910 (see Table \ref{tab:image_para} for image properties). To obtain a high-resolution spectral index map, we select baselines longer than $100\,k\lambda$ and make images of the cluster field with Briggs' \texttt{robust=}$-0.25$ scheme. When imaging the uGMRT data, we also applied a taper with Gaussians (i.e., \texttt{outertaper=}15~arcsec) to obtain a resolution close to that of the LOFAR image (i.e.,$14.7\,{\rm arcsec}\times6.9\,{\rm arcsec}$). The LOFAR and uGMRT primary beam corrected images are smoothed by a 2D Gaussian function to a common resolution of 15~arcsec, corrected for the misalignment between the images using compact sources, and regridded to a common pixel size. These final images were used to calculate the spectral index\footnote{The radio synchrotron spectrum of the form $S\propto\nu^\alpha$ is used in this paper.} as:

\begin{equation}
\alpha=\frac{\log{(S_{\rm LOFAR}/S_{\rm uGMRT})}}{\log{(\nu_{\rm LOFAR}/ \nu_{\rm uGMRT})}}, 
\label{eqn:spx}
\end{equation}
where $\nu_{\rm LOFAR}=145$~MHz and $\nu_{\rm uGMRT}=383$~MHz are the central frequencies of the LOFAR and uGMRT images, respectively. The spectral index error is calculated by summing in quadrature the flux-scale uncertainty associated with calibration (i.e., 10\%\ and 5\% for the LOFAR and uGMRT data, respectively) and the image noise.

We also make a low-resolution spectral index map with a resolution of 22~arcsec following a similar procedure to that described above. We use a common Briggs' \texttt{robust=}$0.25$ weighting for the datasets, an \texttt{outertaper}=20~arcsec for the uGMRT data, and smooth final images to 22~arcsec before the calculation of the spectral indices (via Equation \ref{eqn:spx}).

\begin{table*}
\centering
        \begin{tabular}{lccccc}
                \hline\hline
                Data          &   UV-range     &   $\mathtt{Robust}^a$   & $\theta_{\rm \tiny FWHM}$   &     $\sigma$        & Figure  \\
                & (k$\lambda$) & (\texttt{outertaper}) & (arcsec$\times$arcsec, $PA^b$) & ($\mu{\rm Jy}\,{\rm beam}^{-1}$) & \\ \hline
                LOFAR 145 MHz &   $0.02-41.7$   &  $-0.25$              &   $14.7\times7.0$ ($-85^\circ$)      & 265 & \ref{fig:radio}\\
                      &   $0.02-41.7$   &  $0.25$               &         $19.8\times16.2$ ($66^\circ$)& 340 & \ref{fig:radio}\\ 
                      &   $0.1-41.7$   &  $-0.25$               &         $15.0\times15.0^c$           & 350 & \ref{fig:spx}$^d$, \ref{fig:region_inj}\\    
                      &   $0.1-41.7$   &  $0.25$               &         $22.0\times22.0^c$           & 440 & \ref{fig:spx}$^d$\\    \hline 
                uGMRT 383 MHz &   $0.07-42.3$   &  $-0.25$              &     $6.9\times5.1$ ($57^\circ$)      & 45  & \ref{fig:radio}, \ref{fig:gmrt_hsc} \\ 
                      &   $0.07-42.3$   &  $0.00$ ($15\arcsec$) &         $15.5\times13.9$ ($75^\circ$)& 70  & \ref{fig:radio}\\ 
                      &   $0.1-42.3$   &  $-0.25$ ($15\arcsec$) &         $15.0\times15.0^c$           & 75  & \ref{fig:spx}$^d$, \ref{fig:region_inj}\\      
                      &   $0.1-42.3$   &  $0.25$ ($20\arcsec$) &         $22.0\times22.0^c$           & 125  & \ref{fig:spx}$^d$\\      
                \hline\hline
        \end{tabular}\\ 
        Notes: $^a$: Briggs weighting of the visibility; $^b$: position angle; $^c$: the image is smoothed; $^d$: spectral index map
        \label{tab:image_para}
        \caption{Radio image properties.}
\end{table*}

\section{A supercluster discovered in the eFEDS field}
\label{sec:supercluster_definition}

In the literature, a variety of definitions are used to characterize a supercluster. The first X-ray flux-limited supercluster catalog by \citep{Chon+13} is built using a definition according to which a supercluster is a group of two or more clusters separated by a distance of less than a linking length. The linking length depends on the cluster number over-density threshold $f$ \citep[as in ][]{zucca+93}, and the survey area used in the search. Therefore, the supercluster definition is highly dependent on the redshift because the over-density is computed with respect to the comoving mean cluster density \citep[see Eqs.~1, 2, and 3 in ][]{Chon+13}. However, in their later work, \cite{Chon+15} defined superclusters as the structures that will detach from the Hubble flow in the future, and thus collapse and form a virialized object. One important result of this new definition is that some of the most famous superclusters in the literature (Shapley, Virgo, and Laniakea) are no longer  considered as such because most of their matter content will not collapse in the future and are excluded in their latest \cite{Chon+15} supercluster catalog.
In this work, we adopt the previous definition of a supercluster \citep{Chon+13} with a cluster number over-density of $f = 10$, and assuming that number of members adds up to five clusters or more. The linking length on the other hand is calculated using the dedicated eROSITA simulations \citep{Comparat+2020}. 
{ More specifically, the simulation allows the number of clusters in redshift bins to be counted, which, when dividing by the volume, gives the cluster number density in each bin. The linking length is just the inverse cube root of $f$ times the cluster number density. }
We find that the physical linking length is 24.6~Mpc (or 79~arcmin) at the redshift of the 
supercluster: $z = 0.36$.

A systematic analysis of the spatial position of the confirmed clusters in the eFEDS field combined with their photometric redshifts allowed us to detect { several} superclusters in the eFEDS field. 
{ In this work we focus on the supercluster at a redshift of $\sim$0.36 consisting of eight member clusters, of which the one with the highest redshift has more than five members. The other superclusters will be presented in a future work (Liu et al. in prep).}

{ This supercluster consists of a chain} of eight clusters at similar redshifts ($z = 0.36$, see Table~\ref{tab:cl_list}), located at the same large-scale structure. This system of clusters stretches for about $\sim$1.5~degrees from the north to the south on the plane of the sky, or 27~Mpc considering the physical distance between the furthest clusters in the chain.
To further confirm the detection, we also examined the HSC galaxy density map  (see the details of analysis above) extracted in the same area. The weight of the galaxies is identical to the color filter described in \cite{Klein+19} entering the cluster richness estimate. The weights efficiently{ filter out} galaxies that are not at the cluster redshift, the effective redshift kernel is peaked at $z=0.36$ and the 15 and 84 percentiles are at $z=0.335$ and $z=0.42$. The redshift filtered galaxy density map can be seen in the right panel of Figure~\ref{fig1}.
We observe the striking overlap between the galaxy overdensity at this redshift and the positions of the clusters detected by eROSITA in Figure~\ref{fig1}. This  highlights and confirms the great potential of eROSITA in detecting large-scale structure, enabling statistical studies of its members located at the nodes of the cosmic web.

\begin{table}[t]
\centering
\renewcommand{\arraystretch}{1.5}
\begin{tabular}{ c c c c}
\hline\hline
Cluster & R.A. & Dec. & Redshift \\
\hline
eFEDS~J093501.1+005418 & 143.7546 & 0.9053 & 0.385 \\
eFEDS~J093510.7+004910 & 143.7945 & 0.8195 & 0.367  \\
eFEDS~J093513.3+004746 & 143.8055 & 0.7963 & 0.367 \\
eFEDS~J093520.8+003445 & 143.8368 & 0.5793 & 0.362 \\
eFEDS~J093612.6+001651 & 144.0527 & 0.2811 & 0.367 \\
eFEDS~J093546.4-000115 & 143.9435 & -0.0209 & 0.345 \\
eFEDS~J093543.9-000334 & 143.9330 & -0.0597 & 0.350 \\
eFEDS~J093432.2-002303 & 143.6343 & -0.3844 & 0.346 \\
\hline\hline
\end{tabular}
\vspace{2mm}
\caption{Members of the supercluster detected by eROSITA and HSC from north to south. We provide cluster names, their position in the sky, and photometric redshifts.}
\label{tab:cl_list}
\end{table}
\section{X-ray and optical properties of the supercluster}
\label{sec3}

In this section, we provide a detailed analysis of the eROSITA X-ray and optical data and examine the physical properties of the member clusters of galaxies, and the emission in the interconnecting bridge regions between them.

\subsection{Integrated properties of the members of the supercluster}

\begin{figure}
\includegraphics[width=0.49\textwidth]{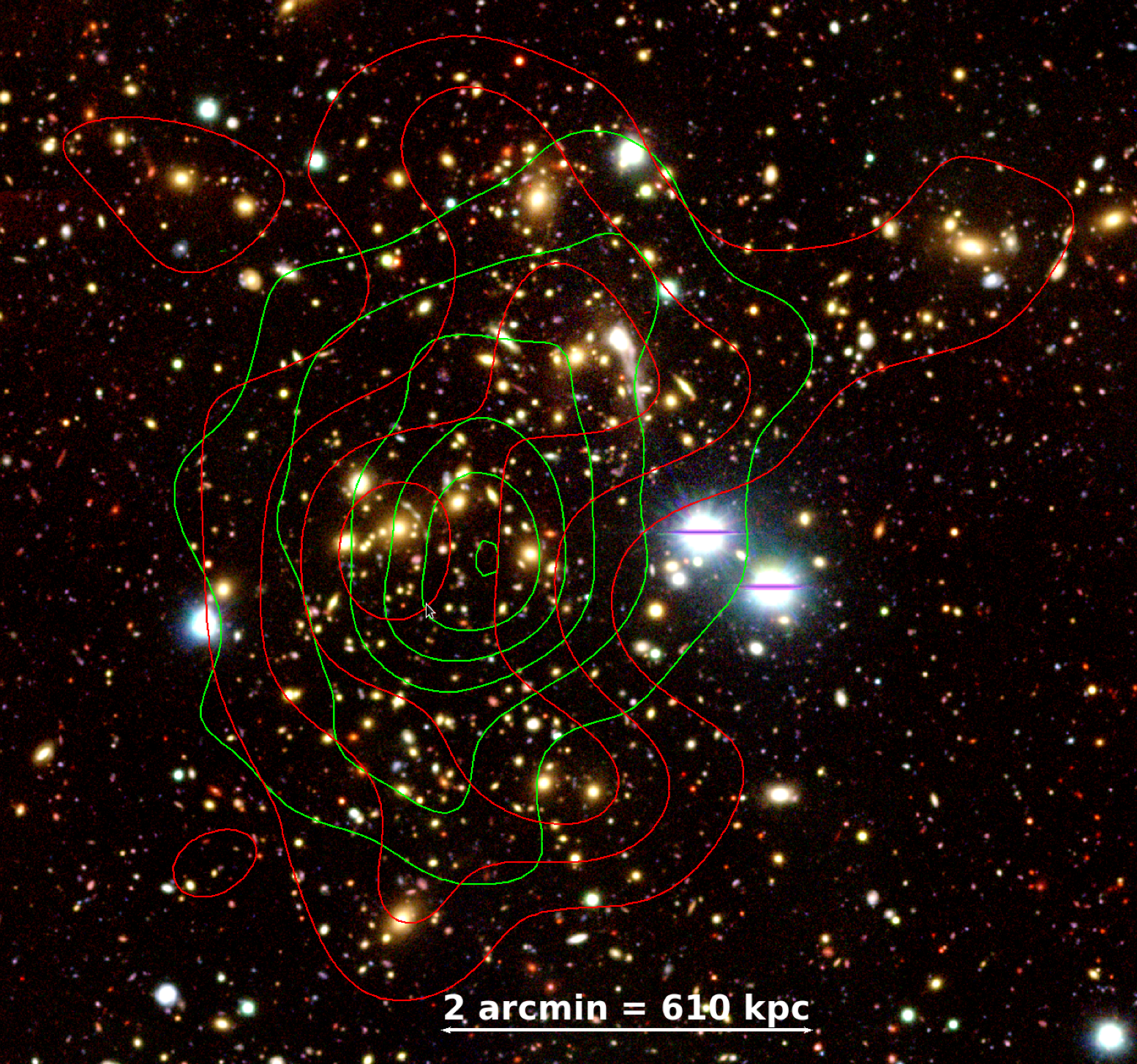}~
\caption{HSC g, r, z color composite image of the cluster system eFEDS~J093513.3+004746 and eFEDS~J093510.7+004910. Galaxy density contours are shown in red and X-ray contours in green. 
} 
\label{fig_opt_0935.1}
\end{figure}

Located at the northern part of the supercluster, eFEDS~J093513.3+004746 is the most massive and luminous cluster in the supercluster, and one of the most massive and luminous in the entire eFEDS field. This cluster lies very close to two  other, fainter clusters, at about 1.5~arcmin  from eFEDS~J093510.7+004910, and 7.1~arcmin from eFEDS~J093501.1+005418. eFEDS~J093513.3+004746 was also detected in the previous optical and SZ survey catalogs, for example SDSS \citep{SDSS_catalog}, redMaPPer \citep{Rozo+15}, and \emph{Planck} \citep{PSZ2}. In the optical data, eFEDS~J093513.3+004746 appears to be very complex with multiple cores and no obvious dominating brightest cluster galaxy (BCG), as seen in Figure~\ref{fig_opt_0935.1}. The richness ratio of these clusters (see Table~\ref{tab:properties}) is approximately 3:2:1. This would indicate that eFEDS~J093513.3+004746 and eFEDS~J093510.7+004910 are undergoing a major merger, while eFEDS~J093501.1+005418, being further away with slightly overlapping $R_{500}$ is likely a pre-merging cluster. The remaining clusters,  eFEDS~J093520.8+003445, eFEDS~J093612.6+001651, eFEDS~J093546.4-000115, eFEDS~J093543.9-000334, and eFEDS~J093432.2-002303, do not show any X-ray or optical atypical properties.

The integrated X-ray properties, for example gas mass, total mass, and gas mass fraction of the clusters within R$_{500}${, obtained using the scaling relation of \cite{bulbul19}  between core excised luminosity and mass,} are given in Table~\ref{tab:properties}. In particular, for eFEDS~J093513.3+004746, the total mass calculated using the \cite{bulbul19} $L_{X}-M_{500}$ scaling relations ($6.0_{-0.5}^{+0.3}\ \times 10^{14}$~M$_{\odot}$) is fully consistent with the halo mass of $5.5_{-0.7}^{+0.6}\cdot10^{14}M_{\odot}$ reported in the Planck catalog \citep{planck+16} at 1$\sigma$ confidence level. 
We remind the reader that the relation between X-ray luminosity and \emph{Planck} Compton $y$-parameter is quite scattered \citep{planck+11}, and therefore the agreement between the two recovered masses is not a conclusive measurement of the mass of the system.

\begin{table*}[t]
\centering
\renewcommand{\arraystretch}{1.5}
\begin{adjustbox}{max width=\textwidth}
\begin{tabular}{ c c c c c c c c c c c }
\hline\hline
Cluster  & $\lambda$ & $T_{cin}$ & $L_{cin}$ & $T_{cex}$ & $L_{cex}$ & $R_{500}$ & $M_{500}$ & $M_{g,500}$ & Counts \\
 & & [keV] & [$10^{44}$ erg/s] &  [keV] & [$10^{44}$ erg/s] & [arcmin] & [$10^{14} M_\odot$] & [$10^{14} M_\odot$]\\
 \hline
 eFEDS~J093501.1+005418  & $143 \pm 12$ & $4.4_{-1.4}^{+4.7}$ & $0.96_{-0.08}^{+0.08}$ & $3.8_{-1.2}^{+2.9}$ & $0.63_{-0.04}^{+0.07}$ & $2.68_{-0.04}^{+0.06}$ & $2.5_{-0.5}^{+0.5}$& $0.21_{-0.02}^{+0.01}$ & $276$ \\
eFEDS~J093510.7+004910  & $62 \pm 9$ & $9.0_{-4.0}^{+10.4}$ & $0.38_{-0.21}^{+0.26}$ & $7.0_{-2.8}^{+8.5}$ & $0.30_{-0.19}^{+0.23}$ & $2.35_{-0.45}^{+0.31}$ & $1.6_{-0.8}^{+0.8}$& $0.11_{-0.06}^{+0.07}$ & $125$ \\
eFEDS~J093513.3+004746  & $208 \pm 15$ & $4.2_{-1.0}^{+1.8}$ & $2.92_{-0.33}^{+0.27}$ & $4.1_{-0.9}^{+1.7}$ & $2.25_{-0.27}^{+0.23}$ & $3.67_{-0.10}^{+0.08}$ & $5.8_{-1.1}^{+1.1}$& $0.61_{-0.05}^{+0.04}$ & $935$ \\
eFEDS~J093520.8+003445  & $30 \pm 6$ & $2.0_{-0.7}^{+2.9}$ & $0.31_{-0.04}^{+0.04}$ & $3.9_{-2.0}^{+8.4}$ & $0.24_{-0.04}^{+0.05}$ & $2.30_{-0.10}^{+0.10}$ & $1.4_{-0.3}^{+0.3}$& $0.10_{-0.01}^{+0.01}$ & $106$ \\
eFEDS~J093612.6+001651  & $32 \pm 7$ & $2.3_{-0.8}^{+12.7}$ & $0.41_{-0.06}^{+0.06}$ & $2.2_{-0.7}^{+7.3}$ & $0.33_{-0.06}^{+0.07}$ & $2.40_{-0.10}^{+0.10}$ & $1.7_{-0.4}^{+0.4}$& $0.13_{-0.01}^{+0.01}$ & $127$ \\
eFEDS~J093546.4-000115  & $41 \pm 7$ & $7.4_{-3.9}^{+12.2}$ & $0.39_{-0.07}^{+0.07}$ & $10.8_{-6.6}^{+14.0}$ & $0.22_{-0.05}^{+0.07}$ & $2.30_{-0.12}^{+0.14}$ & $1.3_{-0.3}^{+0.3}$& $0.10_{-0.02}^{+0.02}$ & $147$ \\
eFEDS~J093543.9-000334  & $41 \pm 7$ & $6.5_{-4.2}^{+12.9}$ & $0.28_{-0.05}^{+0.06}$ & $7.5_{-4.7}^{+11.8}$ & $0.22_{-0.05}^{+0.07}$ & $2.26_{-0.12}^{+0.14}$ & $1.3_{-0.3}^{+0.3}$& $0.10_{-0.02}^{+0.02}$ & $101$ \\
eFEDS~J093432.2-002303  & $58 \pm 8$ & $6.1_{-3.1}^{+11.3}$ & $0.59_{-0.06}^{+0.05}$ & $3.7_{-1.4}^{+4.2}$ & $0.45_{-0.04}^{+0.06}$ & $2.78_{-0.06}^{+0.08}$ & $2.1_{-0.4}^{+0.4}$& $0.18_{-0.01}^{+0.01}$ & $231$ \\
\hline\hline
\end{tabular}
\end{adjustbox}
\vspace{2mm}
\caption{Properties observed by eROSITA and HSC of the eight clusters in the supercluster; see Figure~\ref{fig1}. We show the richness, best-fitting temperatures, and luminosity in units of $keV$ and $10^{44}$ erg/s, respectively, both core included and excluded, the masses and the gas masses both in units of $10^{14} M_\odot${ and the cluster counts in the [0.5-2.0] keV energy band. We remind the reader that the masses are computed using \cite{bulbul19} scaling relations, and the uncertainties reported refer to the sum in quadrature of the statistical uncertainty and the intrinsic scatter present in the scaling relations.   } }
\label{tab:properties}
\end{table*}
\subsection{X-ray dynamical properties of the members of the supercluster}
We are able to measure the density profiles and average temperatures of the member clusters out to $R_{500}$ from the eROSITA data alone. These measurements allow us to determine the morphological parameters of these clusters, which can be directly compared with previous results in the literature of differently selected cluster samples and the clusters detected in the eFEDS field  \citep[e.g.,][]{lovisari17,rossetti17,santos+17}. 

We provide  estimates of the concentration \citep[$c_{SB}$,][]{santos+08}, cuspiness \citep[$\alpha$,][]{Vikhlinin+07}, central density \citep[$c_{SB}$,][]{hudson+10}, and ellipticity ($\epsilon$) parameters of the members of the supercluster based on the eROSITA X-ray data in Table~\ref{tab:morph}. We find that the morphological properties of the cluster members 
do not show any significative deviation from the distribution for all morphological parameters of the clusters detected in the eFEDS field (Ghirardini et al. in prep.); see Fig.~\ref{fig_morph}. We conclude that there is no statistically significant difference between the dynamical state and morphological quantities of clusters located in this supercluster and those of the general eFEDS cluster population.

\subsection{Surface brightness profiles in sectors}
\label{sec:sb_profile}
In the hierarchical structure-formation process, clusters of galaxies grow by accretion of material along the filaments and by merging activities (e.g., Kravtsov \& Borgani 2012, for a review). 
Cluster mergers transform their energy from kinetic motion to plasma thermal and nonthermal (turbulence, bulk motion, and particle acceleration) energy by leaving sharp features on the X-ray measured quantities as well as potentially giving rise to diffuse radio emission, allowing us to study the assembly and growth processes of clusters of galaxies. 
Shocks are marked as pressure discontinuities, where the  gas temperature and density are lower in the pre-shock with respect to the post-shock region. At the cold front regions on the other hand, the X-ray surface brightness and gas density drop sharply, while the temperature of the gas rises sharply in the post-cold-front region, leaving the pressure across the cold front continuous \citep[see][for a review]{Markevitch+07}. Shocks  and  cold  fronts  have  been  observed  in  several galaxy clusters that are clearly undergoing significant merging activity \citep[e.g.,][]{Markevitch+02,Markevitch+05,Owers+09,ghizzardi+10,Russell+10,planck_13_merge,Dasadia+16,tholken2018,Botteon+18,2019arXiv191109236O,Walker+20}.

\begin{figure*}[t]
\centering
\raisebox{-0.45\height}{\includegraphics[width=0.35\textwidth]{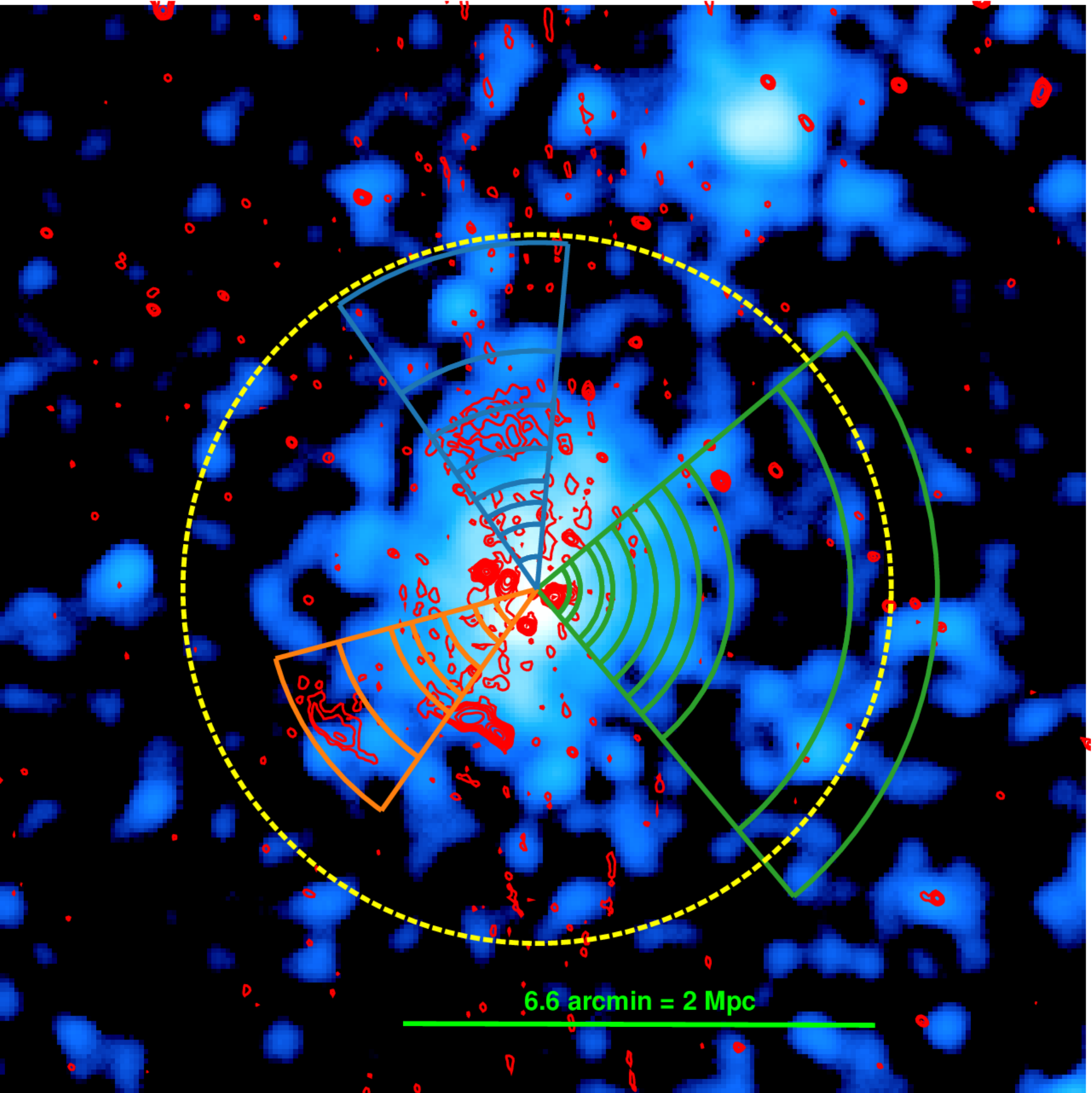}}~
\raisebox{-0.5\height}{\includegraphics[width=0.5\textwidth]{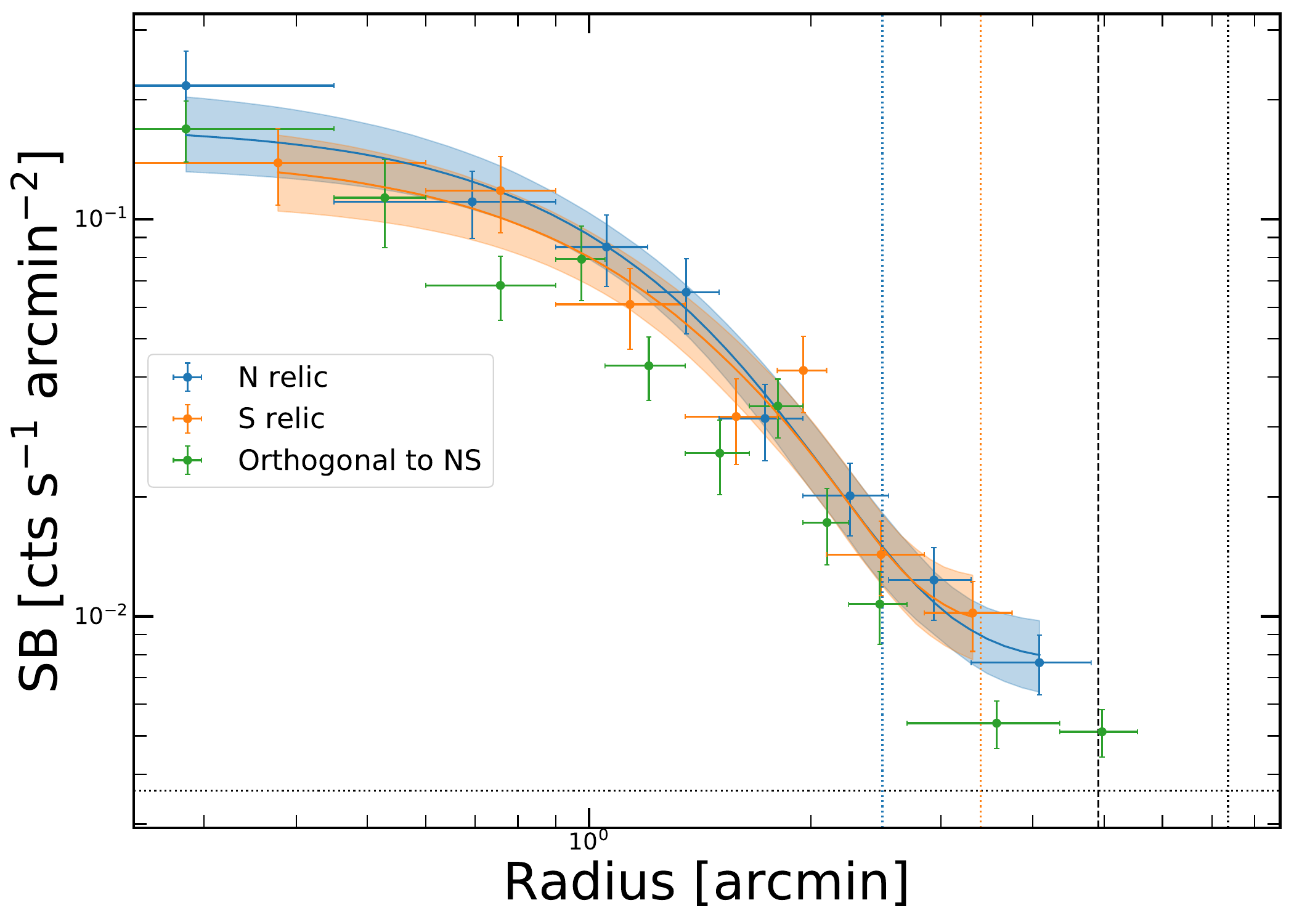}}~
\caption{\emph{Left}: Regions from which we extract the surface brightness profiles. The dashed yellow circle represents the location of $R_{500}$. We show in red the radio contours superimposed on the X-ray image. \emph{Right}: Surface  brightness profiles toward the N relic (in blue), toward the S relic (in orange), and orthogonal to the N--S filament (in green) in the cluster system eFEDS~J093513.3+004746 and eFEDS~J093510.7+004910. The data points represent the observed surface brightness while the solid lines represent the best-fit to the data using the \cite{Rossetti+07} model. The shadowed area represents the 1$\sigma$ distributions around the best fits. The dotted vertical lines represent the location of the N relic (in blue), and of the SE relic (in orange), respectively. } 
\label{fig_SB}
\end{figure*}

We examine the X-ray and radio images of the major merger  detected in this supercluster. The northern region has also recently been targeted for 6.5 hours with uGMRT. The radio data suggest the existence of the two radio relics (2.4~arcmin away from the cluster center in the north and 3.4~arcmin in the south) along the large-scale structure (see Section \ref{sec:radioresults} and Fig.~\ref{fig:spx}). We examine the surface brightness profile of eFEDS~J093513.3+004746 in three wedge regions along and perpendicular to the large-scale structure in the direction of the relics in the X-ray image. Because of the statistical limitations at this depth, we are not able to extract X-ray temperature profiles. We choose the width and the location of our wedge regions to have at least 150~counts in each region. The resulting sectors are displayed in the left panel of Figure~\ref{fig_SB}  . We then fit the surface brightness profiles of the two wedges toward the two relics with a model, given in Equation~\ref{eqn:n_e}, consisting of a beta model and a power-law model, integrated along the line of  sight and convolved by the eROSITA PSF \citep[see][]{Rossetti+07}. The electron number density is

  \begin{equation}
    n_e (r) = n_0
    \begin{cases}
      \left( 1 + \left( \frac{r}{r_c}\right)^2 \right)^{\frac{-3 \beta}{2}}, & \text{if}\ r < r_{J} \\
      \frac{1}{J}\left( 1 + \left( \frac{r_J}{r_c}\right)^2 \right)^{\frac{-3 \beta}{2}} \left( \frac{r}{r_J}\right)^{-\alpha}, & \text{if}\ r > r_{J},
    \end{cases}
    \label{eqn:n_e}
  \end{equation}

\noindent where $n_0$, $r_c$, and $\beta$ are the usual $\beta$-model normalization, core radius, and slope, respectively, $r_J$ indicates the location where the density becomes a power law, $\alpha$ is the slope of the power law, and $J$ measures the density jump at the location of the discontinuity. The fit is performed using the Bayesian nested sampling algorithm MultiNest \citep{multinest}.

We show the surface brightness profile and the best-fit models from the density model in Figure~\ref{fig_SB} (right panel). Interestingly, there is a slight 1$\sigma$ excess of the surface brightness and number density of electrons ($S_{x}\propto \int n_{e}^{2}\, dl$) along the filament direction 
to the south and to the north with respect to the off-filament direction. This could be due to the accretion of clumps and infalling material along the filament direction onto the cluster in the form of hot diffuse gas  \citep[as found also in simulation studies, e.g.,][]{vazza13}. 
Previously, such excess has been observed in clusters very close
to other clusters, for example in the A3391/95 system \citep[][ Reiprich at al. in prep]{Alvarez+18} and A1750 \citep{Bulbul+16}. 
Although there is a hint of a surface brightness jump at 2~arcmin away from the center of  eFEDS~J093513.3+004746 to the south, it is not statistically significant. Because of the statistical limitations of the shallow eFEDS survey, our surface brightness measurements do not reach the southern relic, which is 3.4 arcmin away from the cluster center. Unfortunately, the regions cannot be sufficiently small to detect surface brightness features at the southern and northern relics. Deeper follow-up observations with existing X-ray observatories (e.g., XMM-Newton and Chandra) are required to significantly detect surface brightness and temperature jumps at the locations of the radio relics.

\begin{table*}
\centering
\renewcommand{\arraystretch}{1.5}
\begin{tabular}{ c | c c | c c | c c | c c }
\hline\hline
Region & Density & Gas Mass & Density & Gas Mass & Density & Gas Mass & Density & Gas Mass \\
 & $10^{-4}$ cm$^{-3}$ & $10^{13}M_\odot$ & $10^{-4}$ cm$^{-3}$ & $10^{13}M_\odot$ & $10^{-4}$ cm$^{-3}$ & $M_\odot$ & $10^{-4}$ cm$^{-3}$ & $10^{13}M_\odot$ \\
 \hline
 & \multicolumn{2}{c | }{Z = 0.1 $Z_\odot$, LBG} & \multicolumn{2}{c | }{Z = 0.1 $Z_\odot$, CBG} & \multicolumn{2}{c | }{Z = 0.2 $Z_\odot$, LBG} & \multicolumn{2}{c}{Z = 0.2 $Z_\odot$, CBG} \\
 \hline
Region 1 & 13.7 & 5.2 & 17.3 & 6.5 & 13.6 & 5.1 & 10.3 & 3.9 \\
Region 2 & 3.1 & 10.1 & 3.8 & 12.3 & 2.2 & 7.2 & 2.1 & 6.7 \\
Region 3 & 1.1$\times10^{-4}$ & 2.5$\times10^{-4}$ & 4.1 & 9.8 & 2.9$\times10^{-4}$ & 2.21$\times10^{-4}$ & 3.3 & 7.7 \\
Region 4 & 5.7 & 31.4 & 5.5 & 30.1 & 2.9 & 15.9 & 3.0 & 16.5\\
\hline\hline
\end{tabular}
\vspace{2mm}
\caption{Upper limits on the properties of the intergalactic gas between member clusters of the supercluster for two background regions extracted from the LBG and CBG regions marked in dashed green or magenta boxes in Fig~\ref{fig1}.}
\label{tab:properties_whim}
\end{table*}
\begin{table*}[t]
    \centering
    \begin{tabular}{ c c c c c }
\hline\hline
Name & $n_0$ [ $cm^{-3}$ ] & $c_{SB}$ & $\alpha$ & $\epsilon$ \\
\hline
eFEDS~J093501.1+005418 & $16.2_{-4.3}^{+4.7}$  & $0.25_{-0.03}^{+0.03}$  & $0.78_{-0.22}^{+0.18}$  & $0.90_{-0.11}^{+0.07}$ \\ 
eFEDS~J093510.7+004910 & $6.3_{-2.4}^{+5.4}$  & $0.18_{-0.06}^{+0.13}$  & $0.43_{-0.27}^{+0.41}$  & $0.75_{-0.23}^{+0.18}$ \\ 
eFEDS~J093513.3+004746 & $8.7_{-2.2}^{+2.1}$  & $0.15_{-0.01}^{+0.02}$  & $0.42_{-0.19}^{+0.13}$  & $0.85_{-0.06}^{+0.07}$ \\ 
eFEDS~J093520.8+003445 & $10.6_{-3.9}^{+3.9}$  & $0.18_{-0.04}^{+0.05}$  & $0.89_{-0.23}^{+0.14}$  & $0.79_{-0.20}^{+0.14}$ \\ 
eFEDS~J093612.6+001651 & $9.7_{-4.2}^{+4.4}$  & $0.16_{-0.04}^{+0.06}$  & $0.77_{-0.29}^{+0.17}$  & $0.73_{-0.19}^{+0.18}$ \\ 
eFEDS~J093546.4-000115 & $27.4_{-5.2}^{+6.3}$  & $0.35_{-0.06}^{+0.10}$  & $1.29_{-0.15}^{+0.18}$  & $0.81_{-0.17}^{+0.13}$ \\ 
eFEDS~J093543.9-000334 & $7.3_{-3.5}^{+5.4}$  & $0.18_{-0.05}^{+0.07}$  & $0.65_{-0.35}^{+0.28}$  & $0.77_{-0.25}^{+0.17}$ \\ 
eFEDS~J093432.2-002303 & $7.2_{-3.1}^{+4.8}$  & $0.19_{-0.03}^{+0.04}$  & $0.51_{-0.30}^{+0.30}$  & $0.54_{-0.09}^{+0.12}$ \\ 
\hline\hline
    \end{tabular}
    \caption{Morphological parameters measured using eROSITA of the eight clusters in the supercluster. We list central density $n_0$, surface brightness concentration $c_{SB}$, cuspiness $\alpha$, and ellipticity $\epsilon$.}
    \label{tab:morph}
\end{table*}
\subsection{X-ray properties of the plasma emission in between the clusters}

To study the X-ray properties of filamentary gas located in the interconnecting regions  between member clusters of the supercluster, we extract the spectra in solid rectangular regions marked in green in Figure~\ref{fig1}. We refer to these regions as Regions 1, 2, 3, and 4 from the north to south from here on. We note that these regions are selected beyond $R_{200}$ of each cluster to avoid contaminating the WHIM emission with the emission from cluster gas. The background spectra are extracted from the dashed rectangular regions surrounding the solid green regions and marked as the local background (LBG). To carefully account for the spatial variation in the X-ray background, we also use the dashed magenta region in the northwest corner of the image as test background, and refer to this background as the corner background (CBG).

The fitting procedure is very similar to that described in Section~\ref{sec:spectral}; we report the extra steps we follow in this section  to maximize potential detection of the WHIM emission. Given that the eFEDS data are too shallow to obtain constraints on the plasma temperature, we apply a flat prior in the range from $10^6$~K to $2\times 10^7$~K predicted by simulations \citep[see ][]{vazza+19}. Two different metal abundance values are used in the fits, $0.1 Z_{\odot}$ and $0.2 Z_{\odot}$  \citep{Cen+06,Leccardi+08}. As the absorbed bremsstrahlung emission from this gas is expected to be mostly significant below 1~keV, we extend the lower energy band to 0.3~keV for the TMs with on-chip filter (1, 2, 3, 4, and 6), and to 0.5 keV for the TMs without the on-chip filter (5 and 7). Similarly, we use the Goodman-Weare MCMC code implemented within \texttt{xspec} to marginalize over all free parameters, focusing in particular on the posterior distribution on the normalization of the \texttt{apec} model, which is directly linked to the filament gas density and gas mass.
After the fits are performed, no significant emission is detected. We take the $3\sigma$ upper limit (the 99.7~percentile) on the \texttt{apec} normalizations and convert these upper limits to upper limits on the density of the filament, assuming that the gas in the filament has a cylindrical shape whose axis is in the direction connecting two nearby clusters, which 
can then be translated into upper limits on the gas mass present in the filament. We show our upper limits on the number density and gas mass in Table~\ref{tab:properties_whim}. Our upper limits on electron number densities are less than $10^{-4}$~cm$^{-3}$ in the regions we examine regardless of the background used in the analysis. 

The reported values of WHIM gas density in the interconnecting bridges of clusters of galaxies in the literature are of the order of $10^{-4}$~cm$^{-3}$ and 
gas mass in the range of a few $10^{13}-10^{14}$~M$_{\odot}$  \citep{Werner+08, Bulbul+16, eckert+15, planck+2013}. Our upper limits are consistent with the values reported for A1750, A2744, and A133. In region 3, our limits point toward very low density when we use a local background around the bridge. This is probably caused by the fact that the local background signal is enhanced as it is too close to the putative filament.

\section{Radio emission from the merging clusters eFEDS~J093513.3+004746 and eFEDS~J093510.7+004910}
\label{sec:radioresults}

Galaxy clusters that are undergoing mergers can host radio relics and halos. Radio relics are elongated ($\sim$Mpc), steep-spectrum ($\alpha\lesssim-1$) synchrotron sources that sit preferentially at the edges of galaxy clusters \citep[e.g.,][]{Ensslin1998,VanWeeren2010,Brunetti2014,Pearce2017,2019SSRv..215...16V}. There is a consensus that radio relics are produced by large shock waves but  many aspects, including the cause for the apparently high particle acceleration efficiency, remain  unclear \citep[e.g.,][]{Botteon2016b,Botteon2020,vanWeeren2016,Hoang2017,Hoang18,2019PASJ...71...79O,2019arXiv191109236O,Bruggen2020}. Another type of extended synchrotron radio source, namely halos, is often found in the central region of galaxy clusters. Halos have steep spectra, are unpolarized and extend up to approximately megaparsec scales. They are thought to be generated through merger-induced turbulence that re-accelerates CRs in the ICM to  relativistic speeds. In the presence of $\mu$G magnetic fields, the CR electrons emit  synchrotron radio emission at radio wavelengths \citep[e.g.,][]{Brunetti2001,Brunetti2007a,Miniati2015,Brunetti2016,Pinzke2017a}. 

\begin{figure*}
        \centering
        \includegraphics[width=0.49\textwidth]{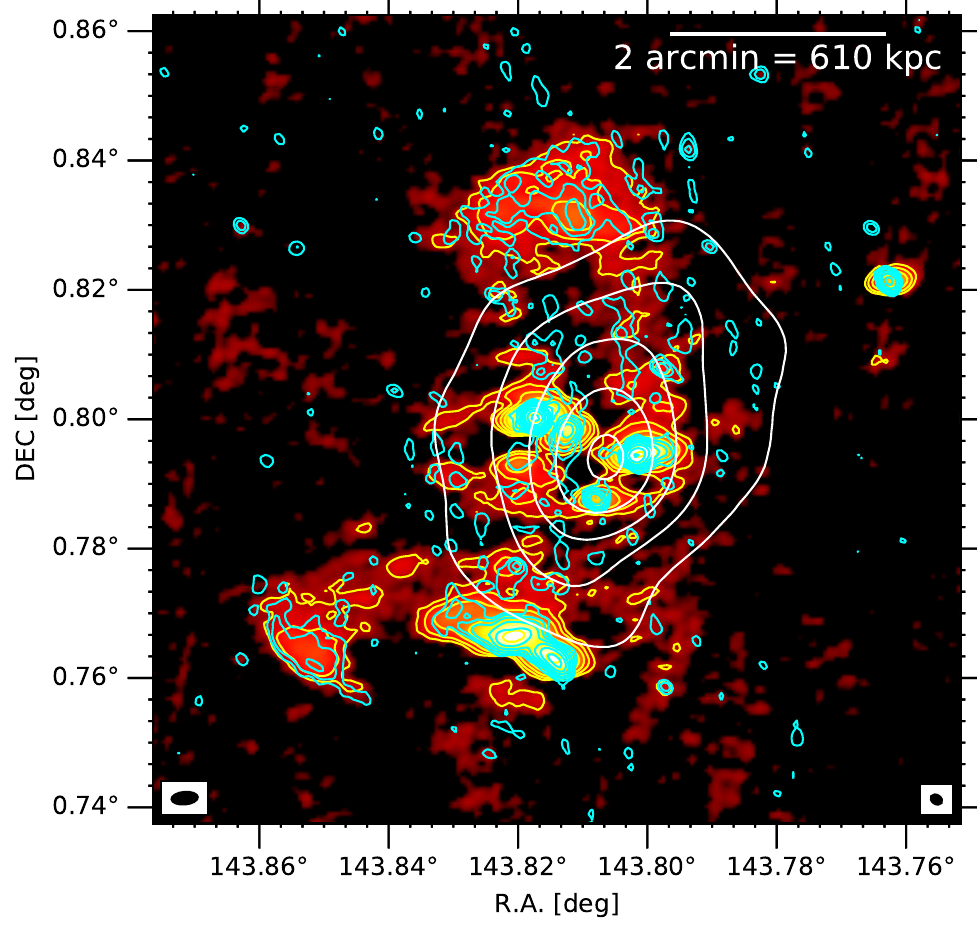} \hfil
        \includegraphics[width=0.49\textwidth]{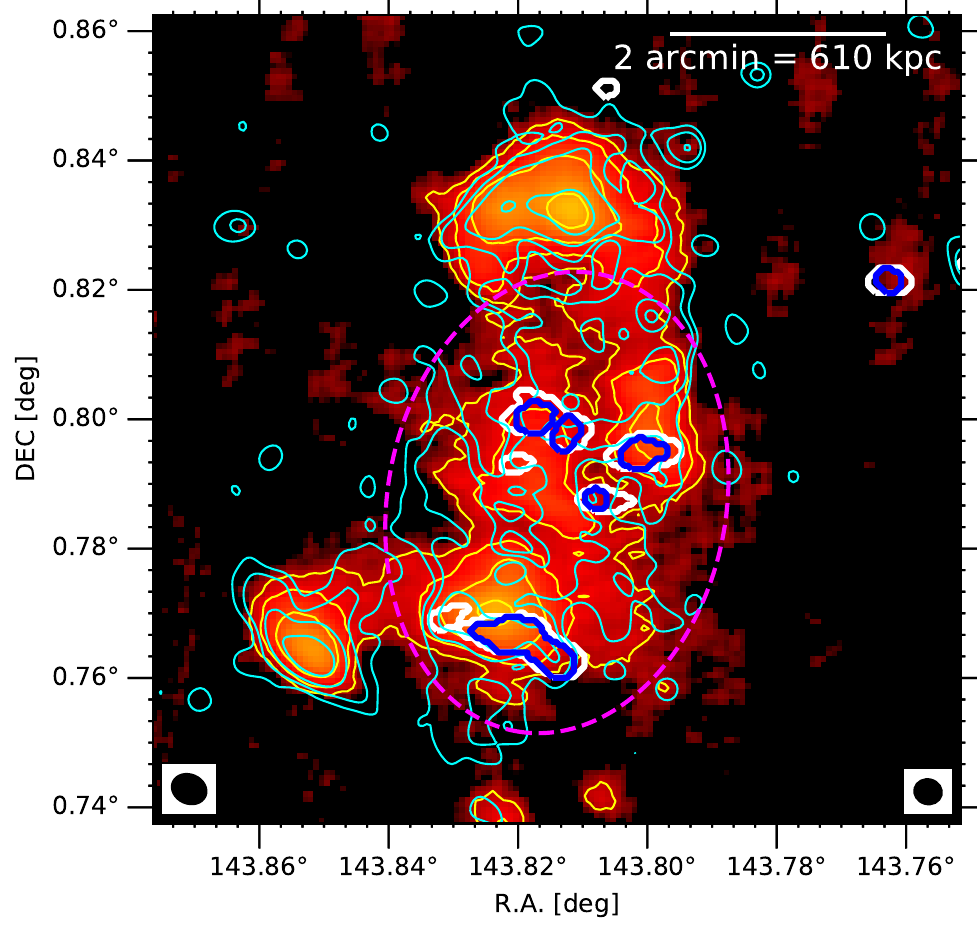}
        \caption{\emph{Left}: LOFAR 145~MHz high-resolution image ($7.5$~arcmin $\times$ $7.5$~arcmin) of the cluster system eFEDS~J093513.3+004746 and eFEDS~J093510.7+004910. The  LOFAR (yellow) and uGMRT (cyan) contours are drawn at levels of $[1,2,4,8,16,32,64]\times3\sigma$ where $\sigma_{\rm LOFAR}=265\,\mu{\rm Jy\,beam}^{-1}$ and $\sigma_{\rm uGMRT}=45\,\mu{\rm Jy\,beam}^{-1}$. The resolution is $14.7\,{\rm arcsec}\times7.0\,{\rm arcsec}$ with a position angle of -85 degree and $6.9\,{\rm arcsec}\times5.1\,{\rm arcsec}$ (${\rm PA}=57\,{\rm degree}$) for the LOFAR and uGMRT images, respectively.  The X-ray eROSITA (white) contours are at the levels of $[0.2, 0.36, 0.52, 0.68, 0.84, 1]~{\rm cts\,s^{-1}\,arcmin^{-2}}$. \emph{Right}: LOFAR 145~MHz and uGMRT 383~MHz low-resolution contour image after the subtraction of point sources which are shown with the white and blue lines, respectively. The LOFAR (yellow) and uGMRT (cyan) contour levels are identical to those in the left image with $\sigma_{\rm LOFAR}=340\,\mu{\rm Jy\,beam}^{-1}$ and $\sigma_{\rm uGMRT}=70\,\mu{\rm Jy\,beam}^{-1}$. Here ${\rm beam_{LOFAR}}=19.8\,{\rm arcsec}\times16.2\,{\rm arcsec}$ (${\rm PA}=66\,{\rm degree}$) and  ${\rm beam_{uGMRT}}=15.5\,{\rm arcsec}\times13.9\,{\rm arcsec}$ (${\rm PA}=75\,{\rm degree}$). The (magenta dashed) ellipse shows the region where the integrated flux density for radio halo is measured. In both images, the LOFAR  and uGMRT synthesized beams are shown in the bottom left and right corners, respectively. }
        \label{fig:radio}
\end{figure*}

\begin{figure}
        \centering
        \includegraphics[width=1\columnwidth]{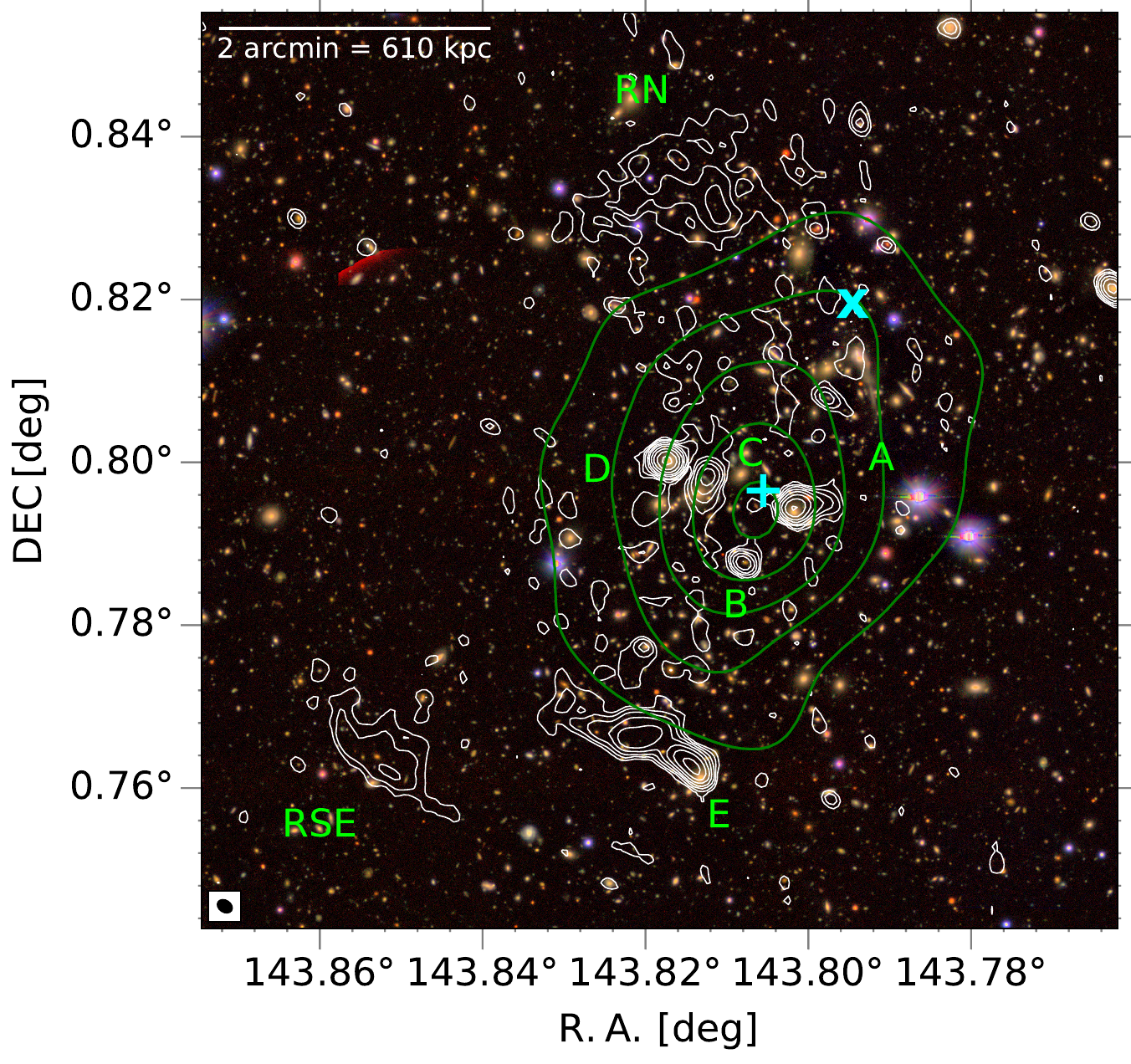}
        \caption{HSC g, r, z color composite image overlaid with the eROSITA (green) and uGMRT (white) contours for the eFEDS~J093513.3+004746 and eFEDS~J093510.7+004910 system. The labels for the radio sources are given. The locations of the sub-clusters eFEDS~J093513.3+004746 and eFEDS~J093510.7+004910 are marked with a plus symbol ($+$) and a cross ($X$), respectively.
        }
        \label{fig:gmrt_hsc}
\end{figure}

\begin{figure*}
        \centering
        \includegraphics[width=0.24\textwidth]{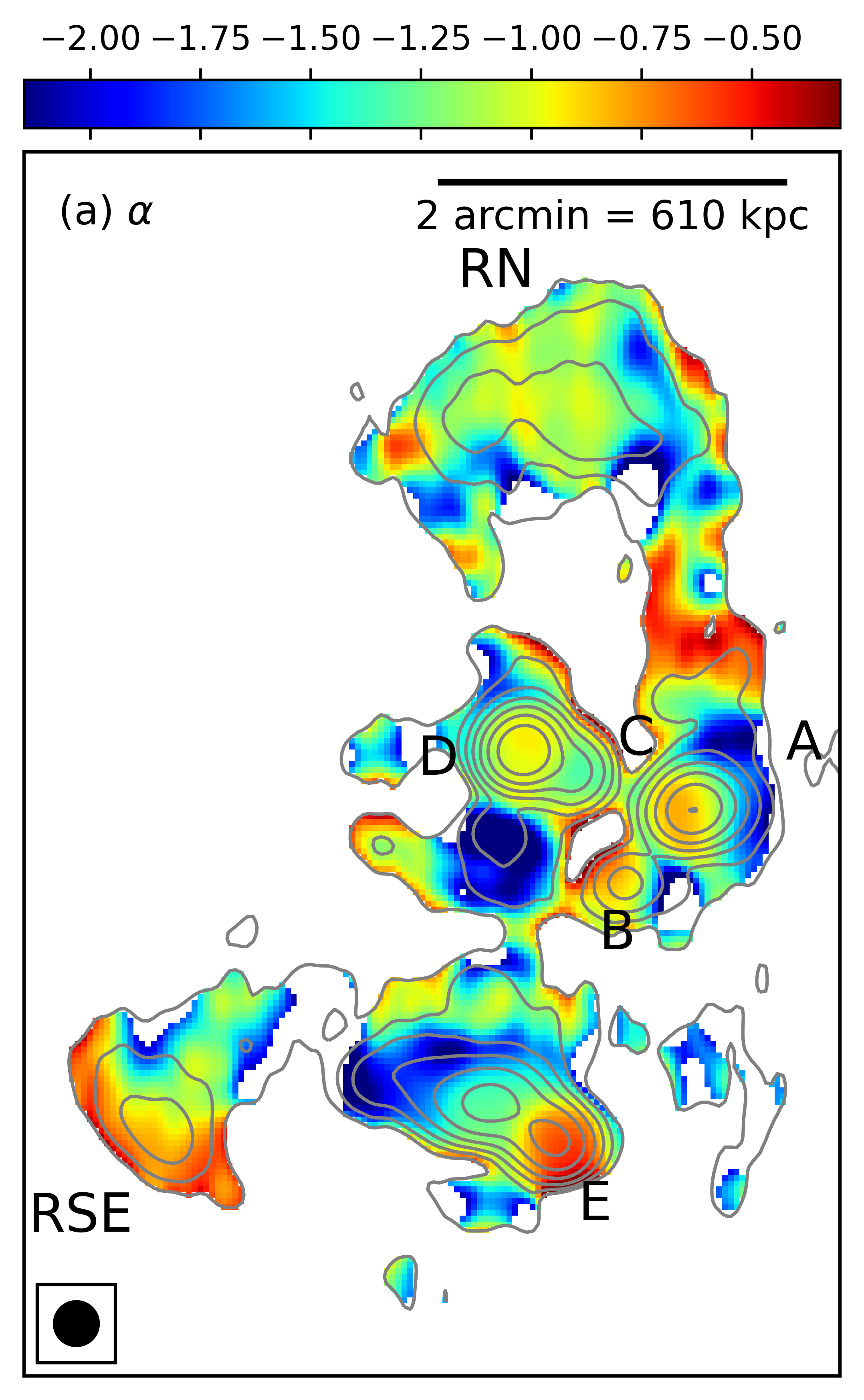} \hfill
        \includegraphics[width=0.24\textwidth]{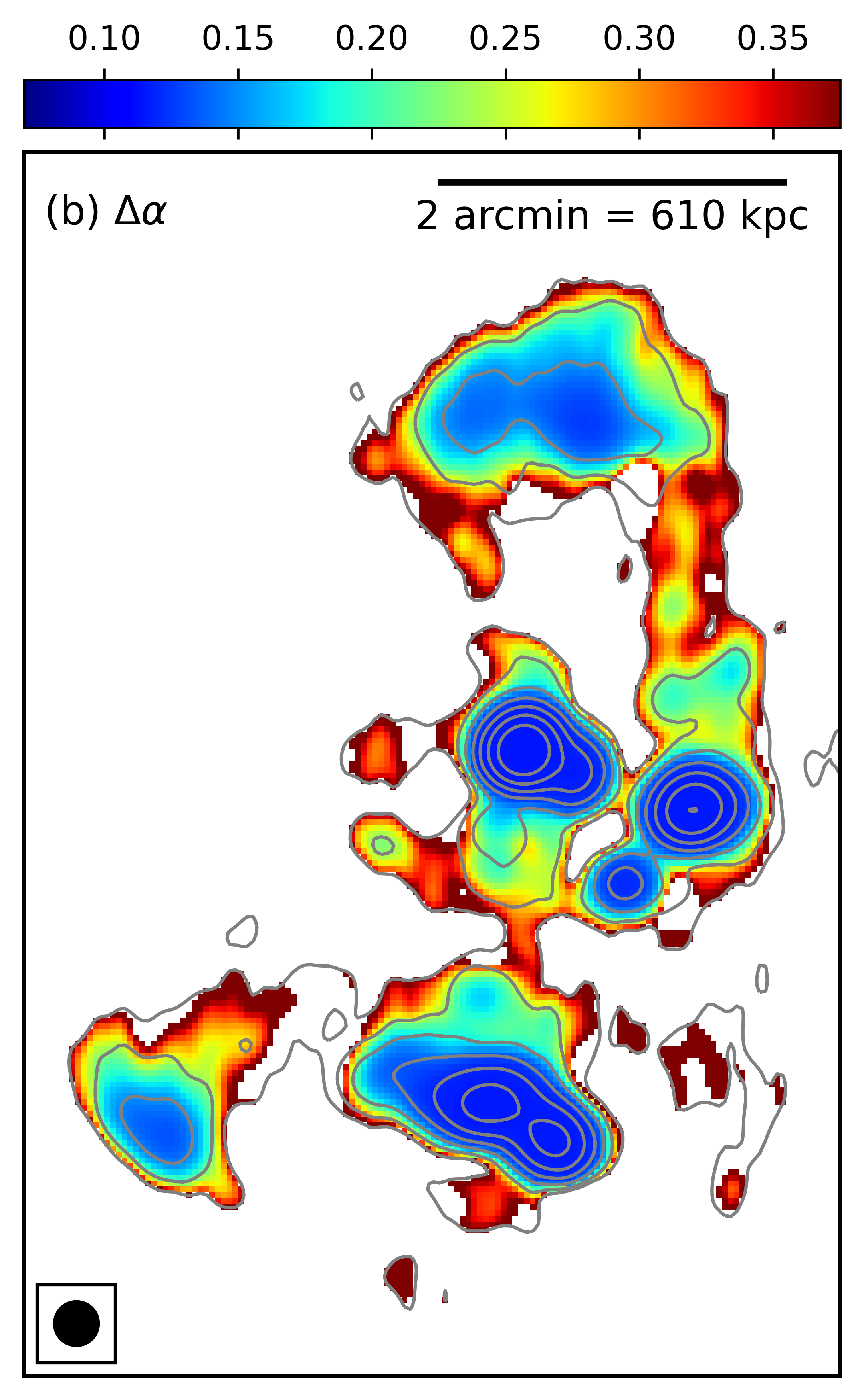}\hfil
        \includegraphics[width=0.24\textwidth]{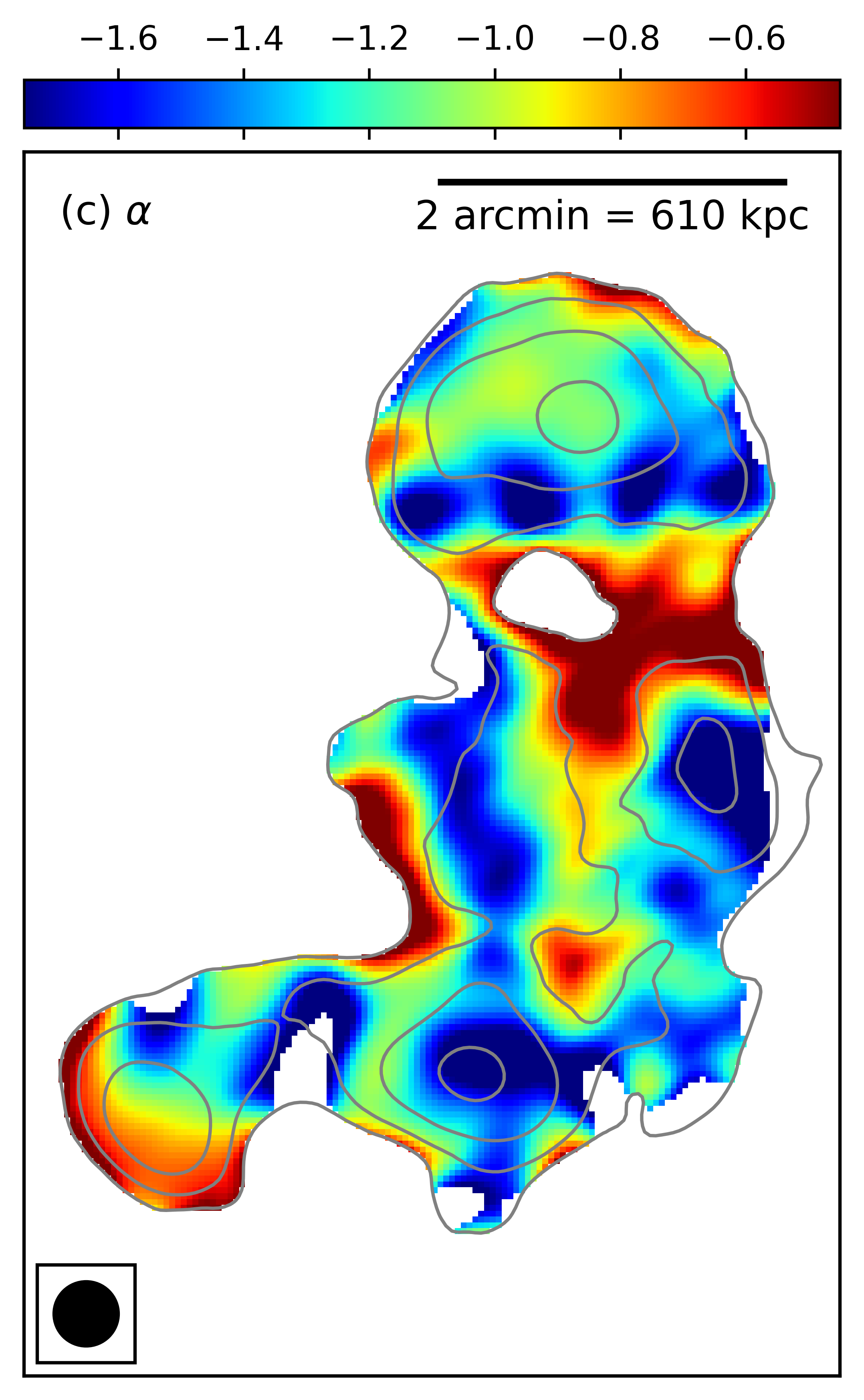} \hfill
        \includegraphics[width=0.24\textwidth]{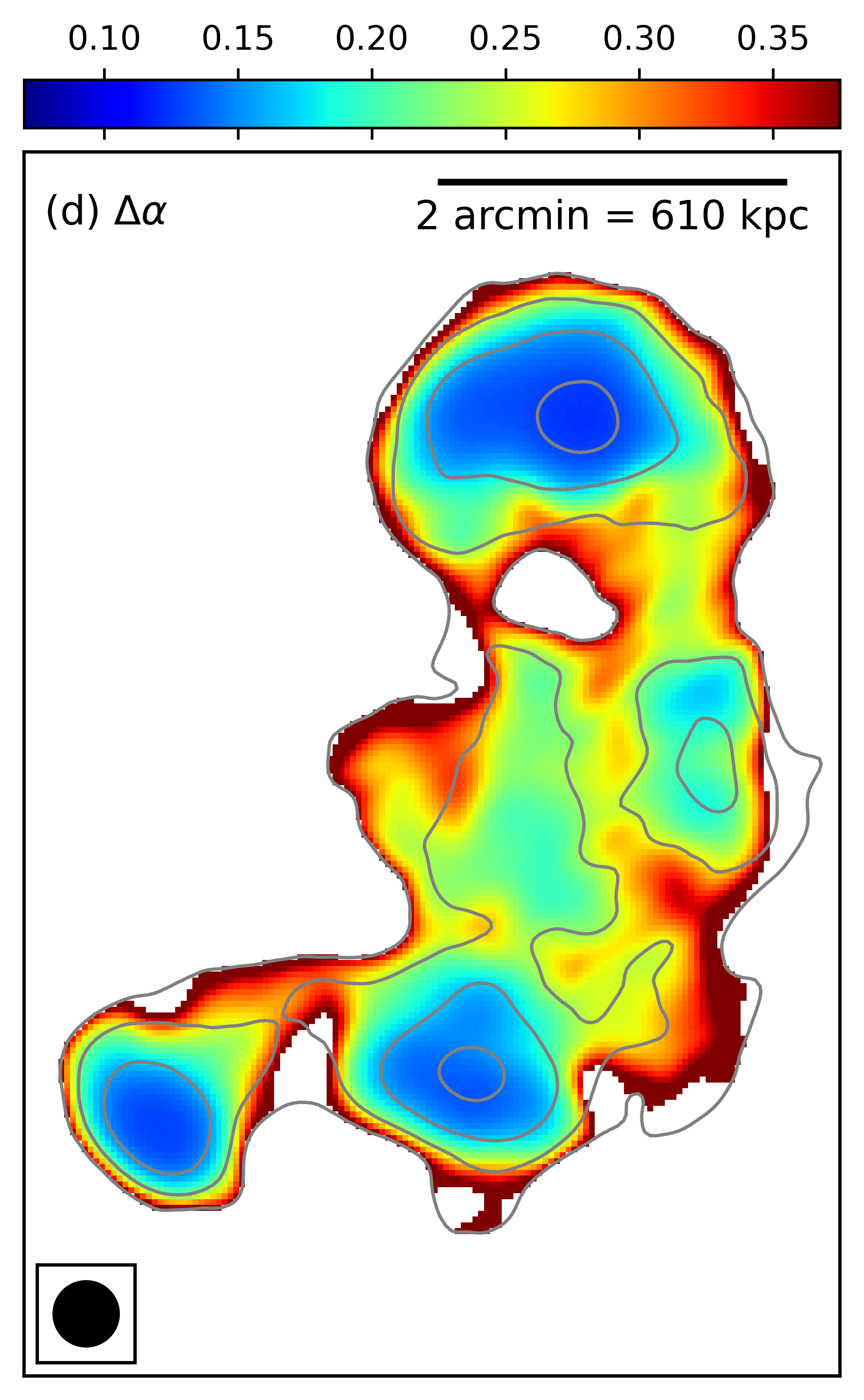}
        
        \caption{Spectral index (a,c) and error maps (b,d) between 145~MHz and 383~MHz of the cluster system eFEDS~J093513.3+004746 and eFEDS~J093510.7+004910 at resolutions of 15~arcsec (a, b) and 22~arcsec (c, d). The LOFAR (gray) contours are drawn at the levels of $[1,2,4,8,16,32]\times3\sigma$ where $\sigma_{\rm LOFAR}=350\,\mu{\rm Jy\,beam}^{-1}$ for the high-resolution images (a, b) and $\sigma_{\rm LOFAR}=440\,\mu{\rm Jy\,beam}^{-1}$ for the low-resolution images (c, d).}
        \label{fig:spx}
\end{figure*}
Figure \ref{fig:radio}  presents the LOFAR 145~MHz and uGMRT 383~MHz images of the cluster system eFEDS~J093513.3+004746 and eFEDS~J093510.7+004910 which is reported to be a single cluster, namely PSZ2 G233.68+36.14, in the \emph{Planck} catalog  \citep{PSZ2}. The images show the detection of extended radio sources in the southeast, central, and northern regions at 145~MHz and 383~MHz. Multiple discrete sources and a tailed radio galaxy are also seen in the cluster central and southern regions. We label the sources in Figure \ref{fig:gmrt_hsc} and present their observational properties in the sections below.

\subsection{Southeast and north extended radio sources}
\label{sec:relics}

The southeast (RSE) and north (RN) extended sources shown in Figure \ref{fig:radio} are located roughly 1~Mpc from the X-ray peak (${\rm R. A.}=143.8061$, ${\rm Dec.}=0.7926$). The projected sizes of RSE and RN are 440~kpc$\times$190~kpc and 570~kpc$\times$380~kpc, respectively. Their major axes are roughly perpendicular to the lines connecting the sources and the X-ray peak. The surface brightness of RSE increases sharply towards its outer edge and gradually decreases towards the center of the cluster system. This trend in surface brightness is not clearly seen in RN where the radio emission is slightly brighter in the central region of the source. Figure \ref{fig:gmrt_hsc} shows that these extended radio sources are not obviously associated with any optical sources. Because of the location, morphological properties, and the lack of optical counterparts for RSE and RN, we classify them as radio relics.
\begin{table*}
        \centering
        \begin{tabular}{lccccc}
                \hline\hline
                Source & $S_{\rm 145\,MHz}$ & $S_{\rm 383\,MHz}$ & $\alpha^{\rm 383\,MHz}_{\rm 145\,MHz}$  &R.A. &  Dec. \\
                &       [mJy]        &       [mJy]        &                                 &  &      \\ \hline
                RSE    &    $26.3\pm2.9$    &    $10.3\pm0.6$    &             $-0.97\pm0.13$   &  143.8520 & 0.7646        \\
                RN     &    $77.6\pm8.0$    &    $21.7\pm1.2$    &             $-1.31\pm0.12$   &  143.8151 & 0.8336         \\
                Halo   &   $144.9\pm29.1$   &   $49.8\pm9.1$    &             $-1.10\pm0.28$    &  143.8148 & 0.7863     \\
                A      &   $99.8\pm10.0$    &    $35.9\pm1.8$    &             $-1.05\pm0.12$   & 143.8013 & 0.7945       \\
                B      &    $14.5\pm1.6$    &    $5.0\pm0.3$     &             $-1.10\pm0.13$   & 143.8077 & 0.7874       \\
                C      &    $33.3\pm3.4$    &    $8.9\pm0.5$     &             $-1.36\pm0.12$   & 143.8123 & 0.7983     \\
                D      &   $169.0\pm16.9$   &    $63.6\pm3.2$    &             $-1.01\pm0.12$   & 143.8173 & 0.8001       \\
                E      &   $185.5\pm18.6$   &    $60.0\pm3.0$    &             $-1.16\pm0.12$   & 143.8136 & 0.7623   \\ \hline\hline
        \end{tabular}\\
        \caption{Flux densities and spectral indices of radio sources in the cluster system eFEDS~J093513.3+004746 and eFEDS~J093510.7+004910.
        }
        \label{tab:sources}
\end{table*} 

We measure the integrated flux densities and spectral indices for the relics and present the measurements in Table \ref{tab:sources}. For the flux density measurements, we select only pixels that are above $3\sigma$  in the high-resolution LOFAR and uGMRT images (Figure \ref{fig:radio}, left). The RN source is roughly two to three times brighter and has a steeper spectral index than RSE (i.e., $-1.31\pm0.12$ for RN compared with $-0.97\pm0.13$ for RSE). The spectral index distribution of radio emission in the cluster vicinity is shown in the low-resolution image in Figure \ref{fig:spx}c. The spectral index for RSE steepens from $\sim-0.80$ in the outer region to $-1.72$ in the inner region of the source. For RN, the spectral index decreases from $\sim-0.95$ to $-1.51$ across the width of the relic towards the cluster center (see Figures \ref{fig:spx}c and \ref{fig:spx_profiles}). Spectral steepening across the width of the relics has been observed in merging galaxy clusters and indicates energy losses due to the inverse-Compton and synchrotron processes in the region behind a shock \citep[e.g.,][]{Giacintucci2008,VanWeeren2010,Hoang2017}. 

In the framework of diffusive shock acceleration, the relativistic electrons that emit synchrotron radiations gain their energy through the Fermi-I process. These electrons are repeatedly reflected across a shock front before being deposited into the downstream region. The spectral index of the injected particles $\alpha_{\rm inj}$ is related to the strength of the shock through the Mach number,

\begin{equation}
\mathcal{M}=\sqrt{\frac{2\alpha_{\rm inj}-3}{2\alpha_{\rm inj}+1} }.  
\label{eqn:M}
\end{equation}
The injected relativistic electrons have a power-law energy distribution of $\nicefrac{dN}{dE}\propto E^{-\delta_{\rm inj}}$ where the spectrum power $\delta_{\rm inj}=2\alpha_{\rm inj}+1$. In the  case of simple planar shocks \citep{Ginzburg1969}, the integrated spectral index, $\alpha_{\rm int}$, that is calculated for the entire relic region is related to the injection index, 

\begin{equation}
\alpha_{\rm inj} = \alpha_{\rm int} + \nicefrac{1}{2},  
\label{eqn:alpha}
\end{equation}
where the fraction ($\nicefrac{1}{2}$) accounts for the spectral steepening of the relativistic electrons due to the inverse-Compton and synchrotron losses in the downstream region of the shock. 

\begin{figure}
        \centering
        \includegraphics[width=1\columnwidth]{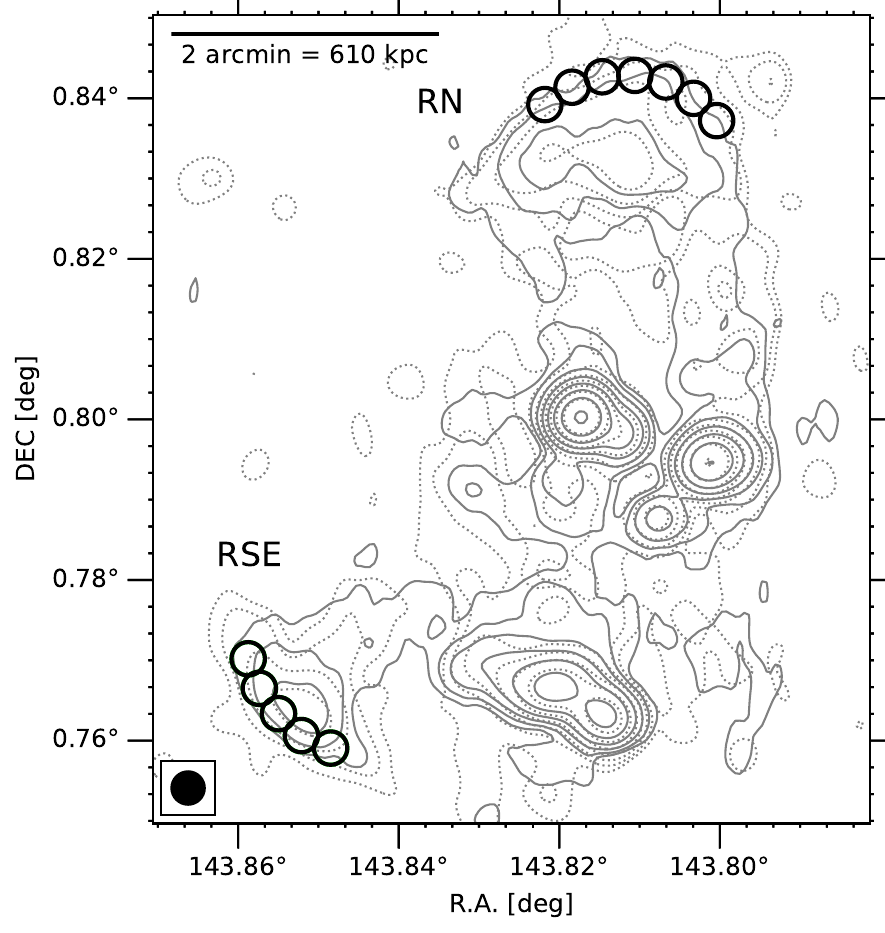}
        \caption{Circular regions where injection spectral indices are extracted in RSE and RN. The diameter of the circular regions is 15~arcsec which is equal to the beam size. The LOFAR (solid) and uGMRT (dotted) contour levels are identical to those in Figure \ref{fig:radio}. Here $\sigma_{\rm LOFAR}=350~\mu{\rm Jy\,beam}^{-1}$ and $\sigma_{\rm uGMRT}=75~\mu{\rm Jy\,beam}^{-1}$.}
        \label{fig:region_inj}
\end{figure}

We use the high-resolution (15~arcsec) images instead of the low-resolution (22~arcsec) images, to measure the injection spectral indices in RSE and RN in order to minimize the mixture of particle populations in the downstream region that have different spectra. The injection spectral indices are $\alpha_{\rm inj}=-0.67\pm0.11$ for RSE and $\alpha_{\rm inj}=-1.07\pm0.11$ for RN. The regions where spectral indices are extracted are shown in Figure \ref{fig:region_inj}. The corresponding Mach numbers for the shocks associated with the relics are $\mathcal{M}_{\rm RSE}=3.5^{+2.3}_{-0.7}$ and $\mathcal{M}_{\rm RN}=2.1^{+0.2}_{-0.1}$. These are typical values for Mach numbers generated through cluster mergers \citep[e.g.,][]{Russell2012,Dasadia2016,Botteon2016a,Botteon2016b}. Radio derived Mach numbers are found to be consistent with those calculated from X-ray data in some cases \citep[e.g.,][]{Macario2013,Shimwell2015,Hoang2017,Hoang2019a,Botteon2020}, but are lower than the X-ray derived Mach numbers in some other cases \citep[e.g.,][]{Govoni2001c,George2015,vanWeeren2016,Pearce2017,George2017,Rajpurohit2018,Hoang2019b,Botteon2020}. In eFEDS~J093513.3+004746 and eFEDS~J093510.7+004910, we were unable to indentify any X-ray surface brightness discontinuity (see Section \ref{sec:sb_profile}).

Alternatively, Mach numbers can be derived from the integrated spectra using Equation \ref{eqn:M} and \ref{eqn:alpha}. The integrated spectral index RN is $-1.31\pm0.12$ which implies a Mach number of $2.7^{+0.7}_{-0.4}$ and is consistent with the value derived from the injection spectral index above. However, this method cannot be applied to the case of RSE because of its flat integrated spectrum. Numerical simulations by \cite{Kang2015a,Kang2015b} indicate that the approximation in Equation \ref{eqn:alpha} should not be used for spherically expanding shocks as the integrated spectra of the relics crucially deviate from the test-particle power laws as the shock decreases its speed over time. Another possibility is that radio relics mark the position of pre-existing clouds of fossil relativistic plasma that are crossed by shocks. In this case, the spectrum of relics generated by weak shocks could reflect the spectrum of the fossil electrons rather than that of Equation \ref{eqn:alpha} \citep[][]{Markevitch+05,Kang2011a}.

\subsection{Central extended radio emission}
\label{sec:halo}
The high-resolution radio image in Figure \ref{fig:radio} (left) shows the presence of extended emission in the central region of the merging system eFEDS~J093513.3+004746 and eFEDS~J093510.7+004910. To enhance the large-scale emission we made a low-resolution image in which compact radio sources are subtracted from the data in Figure \ref{fig:radio} (right). The image shows the presence of extended emission with a projected size of $1.1\,{\rm Mpc}\times 0.750\,{\rm Mpc}$ in the cluster system region connecting the radio relics RSE and RN. The major axis of the extended radio emission is along the same axis as an extension seen in the eROSITA X-ray emission. The large extent of the central emission suggests that it is a radio halo.

We measure the integrated flux density of the radio halo from the point-source subtracted LOFAR and uGMRT images. 
Only pixels that are detected above $3\sigma$ significance within the elliptical region that covers the central region in Figure \ref{fig:radio} are selected. The flux density for the halo is $144.9\pm29.1$~mJy at 145~MHz and $49.8\pm9.1$~mJy at 383~MHz. 
The calculation of the uncertainty is the sum in quadrature of the flux scale error, the image noise (see Section \ref{sec:spx}), and the errors in the subtraction of compact sources (A, B, C, D, and E) that are assumed to be 5\%\ of the total flux density of the compact sources. 
The integrated spectral index of the halo is $-1.10\pm0.28$ which is roughly equal to the estimates for halos in double-relic clusters (i.e., $-1.03\pm0.09$ in the Sausage cluster, \citealt{Hoang2017}; $-1.33\pm0.07$ in MACS~J1752.0+4440, \citealt{Bonafede2012};  $-1.34\pm0.19$ in ClG~0217+70, \citealt{Brown2011a};  $-1.40\pm0.07$ in PSZ1~G108.18-11.53, \citealt{DeGasperin2015}). 

\subsection{Spatial spectral index variations in radio halo}

\begin{figure}
        \centering
                \includegraphics[width=0.35\columnwidth]{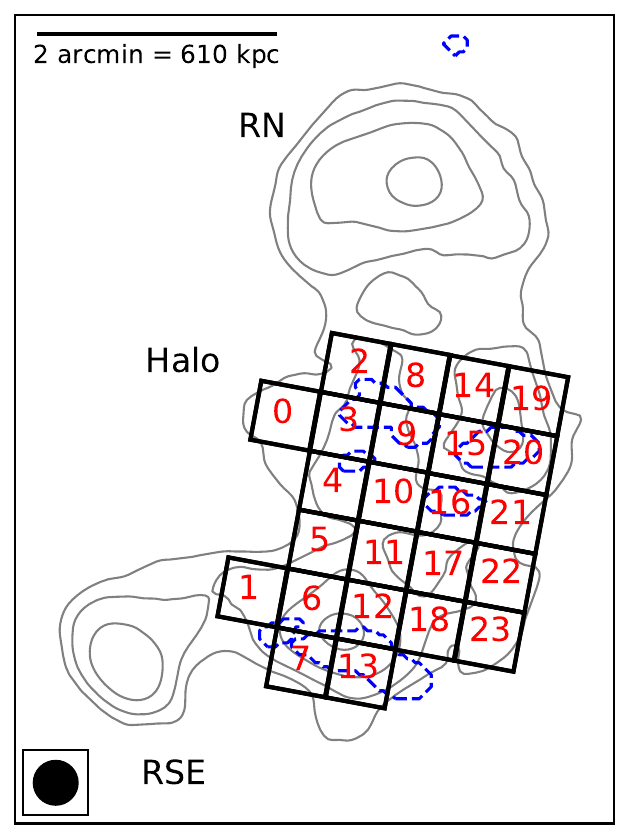}
                \includegraphics[width=0.64\columnwidth]{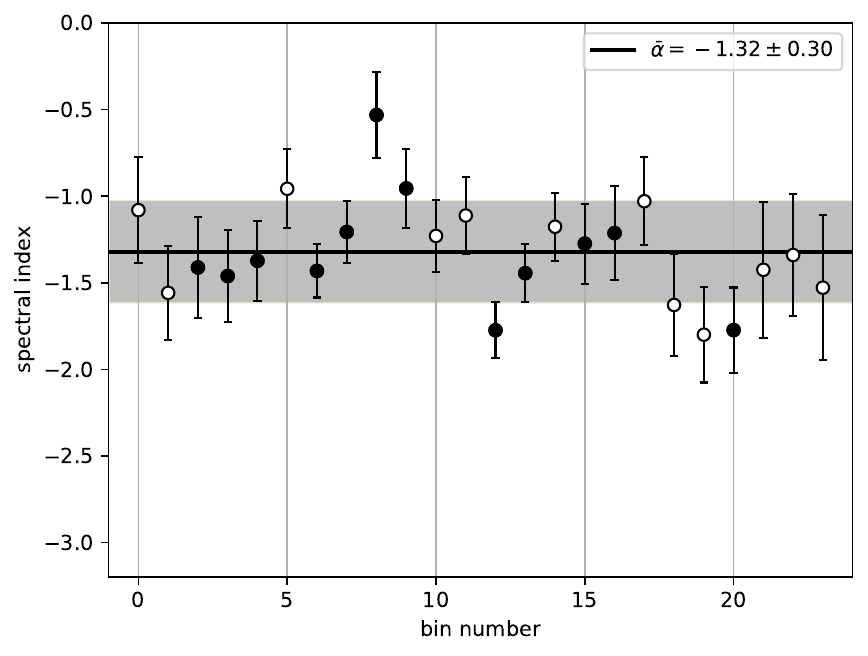}
        \caption{Left: Regions with bin numbers in the cluster system eFEDS~J093513.3+004746 and eFEDS~J093510.7+004910 where spectral indices are extracted. The size of the square regions is 30~arcsec. Right: Spectral indices in the halo region. The horizontal line and the gray region are the mean and $1\sigma$ error for all the data points, respectively. The filled points are the spectral indices in the regions where the compact sources are subtracted (i.e., inside blue dashed contours in the left image).}
        \label{fig:spx_dist}
\end{figure}
The spatial distribution of the spectral indices in the central region of the cluster system is shown in Figure \ref{fig:spx} and is patchy. The spectral index in the northern region of the halo is relatively flat ($\alpha\approx-0.5$), but is steeper in the south. In the halo central region, the spectral indices are also flatter than the regions towards the east and west.

We examine the spectral index distribution in the central region of the cluster system eFEDS~J093513.3+004746 and eFEDS~J093510.7+004910 by extracting spectra in square regions where extended emission is detected with both LOFAR and uGMRT observations. The size of the regions is 30~arcsec which corresponds to a physical scale of 153~kpc at the cluster redshift. We ignore the regions in between the halo and the northern relic because of possible contamination caused by the northern relic. Figure \ref{fig:spx_dist} shows that the spectral indices for 24 regions ($N=24$) over $0.56\,{\rm Mpc}^{2}$ are approximately constant with a mean value of $-1.32\pm0.30$. The mean spectral index of $-1.32$ is steeper than the integrated spectral index of $-1.10$ which is due to the removal of the flat spectrum region from the calculation. We further split the data into two subsets to check if the subtraction of the compact sources has an effect on the spectral index measurement. One set includes the regions where compact sources are subtracted, and the other set is for the regions with the diffuse emission (see Figure \ref{fig:spx_dist}). We find that the mean spectral indices in these two subsets are consistent (i.e. $\alpha_{\rm compact}=-1.32\pm0.34$ and $\alpha_{\rm diffuse} =-1.32\pm0.27$) which implies that the compact-source subtracted regions have similar spectral properties to those without compact sources, which gives us confidence that the removal of the compact sources has not resulted in a systematic error. The intrinsic scatter in projection around the mean is calculated as follows,

\begin{equation}
\sigma_{\rm proj.~intr.}^2 = \sigma^2_{\rm raw} - \sigma^2_{\rm stat.},  
\label{eqn:scatter}
\end{equation}
where $\sigma^2_{\rm raw}=\frac{1}{N-1} \sum_{i=1}^{N}(\alpha_i-\bar{\alpha})^2$
and $\sigma^2_{\rm stat.}=\frac{1}{N}\sum_{i=1}^{N}\sigma_i^2=\frac{1}{N}\sum_{i=1}^{N}\Delta\alpha_i^2$ are the raw and statistical errors, respectively. For the halo region in eFEDS~J093513.3+004746 and eFEDS~J093510.7+004910, we obtain $\sigma_{\rm raw}=0.30$ and $\sigma_{\rm stat.}=0.26$ which results in a projected intrinsic scatter of $\sigma_{\rm proj.~intr.}=0.15$. Following \cite{Hoang2019a}, we assume that the spectral indices in the volume of the halo can vary stochastically around a mean value. The intrinsic scatter of the spectral index on the scale of 153~kpc (i.e., 30~arcsec) is approximated as $\alpha_{\rm intr.}\approx\sqrt{N_{\rm proj.}}\times\alpha_{\rm proj.~intr.}=0.3$, where $N_{\rm proj.}=5$ is the number of cells along the line of sight in the halo volume. Our spectral variations are in line with the measurements in A2256 \citep{VanWeeren2012a}, A665, A2163 \citep{Feretti2004b}, and A2255 \citep{Botteon2020a} in which the spectral variations are suggested to be from inhomogeneous turbulence or due to the inhomogeneous distribution of the mildly relativistic electrons prior to re-acceleration by turbulence. This is different from the cases where spectral indices are also found to be roughly constant across large regions in some other clusters, including the Toothbrush cluster \citep{vanWeeren2016}, the Sausage cluster \citep{Hoang2017} and A520 \citep{Hoang2019a}. 

In the turbulent re-acceleration scenario, large-scale turbulence induced by cluster mergers decays into smaller-scale turbulence that is capable of accelerating mildly relativistic particles. Our measurements of the spectral index variations in the halo would imply turbulent fluctuations on scales that are not much smaller than 150~kpc. 

\subsection{ The scaling relation $P_{\rm 1.4\,GHz}-L_{500}$}

\begin{figure}
        \centering
        \includegraphics[width=1\columnwidth]{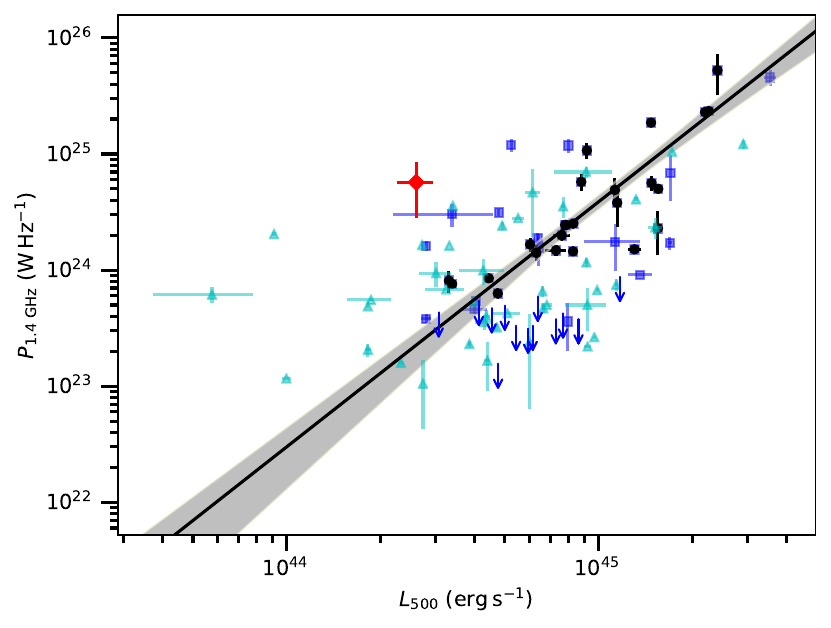}
        \caption{The $P_{\rm 1.4~GHz}-L_{500}$ scaling relation adapted from \protect\cite{Birzan2019} with the data point for the cluster system eFEDS~J093513.3+004746 and eFEDS~J093510.7+004910 in red. The halo samples are taken from \protect\cite[][black circles]{Cassano2013}, \protect\cite[][blue squares]{MartinezAviles2016}, and \protect\cite[][cyan triangles]{Birzan2019}. The upper limits for undetected radio halos are shown with the blue arrows \protect\citep{Cassano2013,Kale2015}. The best-fit relation and the 95\%\ confidence region derived in \protect\cite{Cassano2013}  are also plotted.
        }
        \label{fig:PLx}
\end{figure}

The power of radio halos is known to correlate with the luminosity of their host clusters \citep[e.g.,][]{Cassano2013}. The power for a radio halo at  redshift $z$ can be calculated from the flux density of the source as follows,
\begin{equation}
P_{\rm 1.4\,GHz} = 4\pi k  D_L^2 S_{\rm 1.4\,GHz},  
\label{eqn:power}
\end{equation}
\noindent where $k=(1+z)^{\alpha+1}$  is the $k$-correction term and $D_L$ is the luminosity distance. For the halo in the cluster system eFEDS~J093513.3+004746 and eFEDS~J093510.7+004910, we obtain a power of $P_{\rm 1.4\,GHz}=(5.7\pm2.9)\times10^{24}\,{\rm W\,Hz^{-1}}$. Using the measurement in Table \ref{tab:properties}, we estimate the X-ray luminosity for the cluster system to be $L_{cex} = (2.6^{+0.4}_{-0.3})\times10^{44}$~erg/s.

We compare our measurements for the cluster system eFEDS~J093513.3+004746 and eFEDS~J093510.7+004910 with those for 94 clusters (of which 13 lack a radio
halo detection) presented in \cite{Birzan2019}. Figure  \ref{fig:PLx} shows that our measurements slightly deviate from those for the known radio halos and from the best-fit relation that is obtained for 25 clusters in \cite{Cassano2013}. This deviation might imply that the cluster system is highly disturbed during the merger, which induces turbulence and increases the electron speeds and amplifies magnetic field in the ICM. 

\subsection{Relic--halo connection}
\begin{figure*}
        \centering
        \includegraphics[width=0.38\textwidth, valign=t]{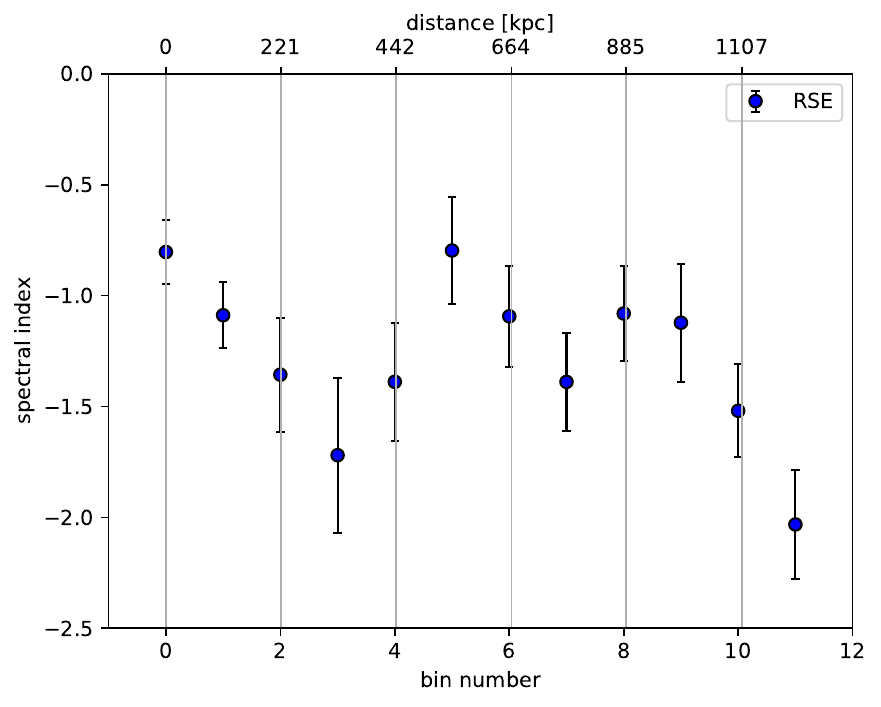} \hfil
        \includegraphics[width=0.205\textwidth, valign=t]{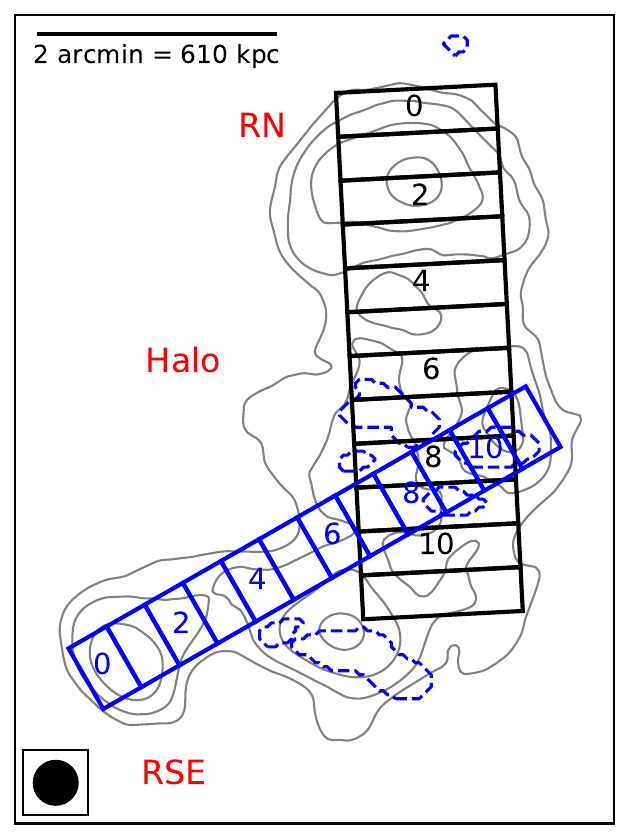} \hfil
        \includegraphics[width=0.38\textwidth, valign=t]{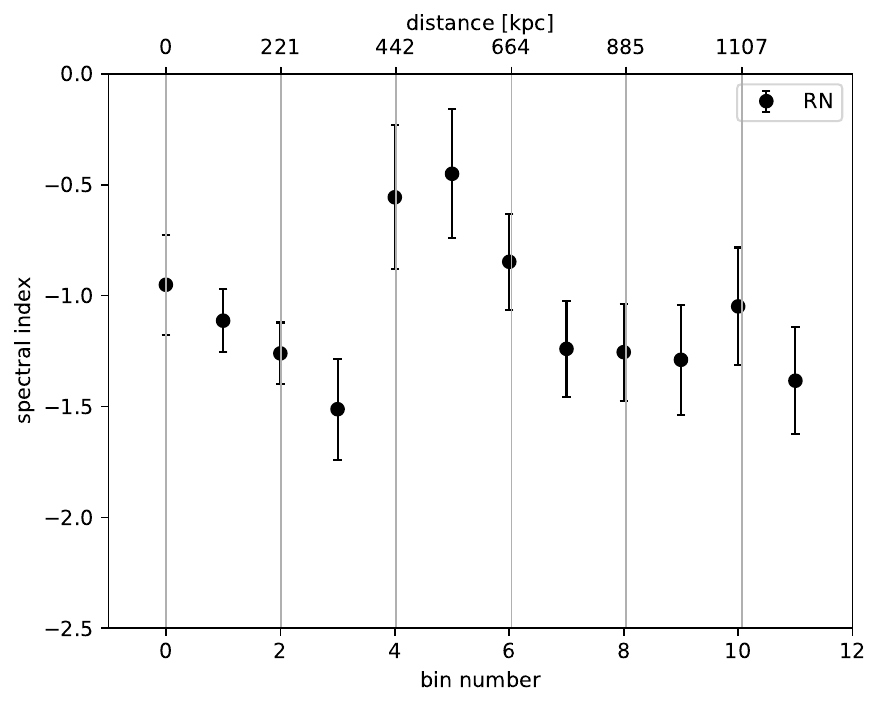}
        
        \caption{Spectral index profiles from RSE (left) and RN (right) towards the cluster center. The middle panel shows the regions where spectral indices are extracted. The width of the regions is equal to the beam size (i.e., 22~arcsec).}
        \label{fig:spx_profiles}
\end{figure*}

Figures \ref{fig:spx} and \ref{fig:spx_profiles} show that the spectral indices steepen across the width of RSE and RN towards the central region of eFEDS~J093513.3+004746 and eFEDS~J093510.7+004910. In the halo region, the spectral indices do not change significantly over a distance of $\sim550$~kpc (i.e., bin numbers from 6 to 10 for RSE and from 6 to 11 for RN). Figure \ref{fig:spx_dist} provides further support to the spectral index variations in the halo. The distinctive spectral behavior in the relics and halo suggests different physical mechanisms for the origin of these extended sources. In the regions between the relics and halo (i.e., bin number 4 for RSE and 4-5 for RN) the spectral indices are flattened. One possibility is that the spectral flattening is generated in a region of short-lived (100~Myr) and intense turbulence that could be generated in the downstream region of the shock connected to RN \citep[e.g.,][]{Markevitch2010}; more robust measurements of the spatial evolution of the spectral index from deeper observations could be used to constrain the strength and decay timescale of this turbulent component.

\subsection{Discrete sources}
\label{sec:ps}
A number of compact radio sources are visible at 145~MHz and 383~MHz in Figures \ref{fig:radio} and \ref{fig:gmrt_hsc}. Most of the notable compact sources A, B, C, and D are located in the central region of the cluster. A tail radio galaxy E is observed at the southern edge of the radio halo. The spectroscopic redshift of source E (i.e., $z=0.350437\pm0.000099$; \citealt{aguado2019fifteenth}) suggests that it is a member of the cluster system. The integrated flux densities and spectral indices of the sources are given in Table \ref{tab:sources}. Their spectral indices range between $-1.01$ (D) and $-1.36$ (C). The source E has a spectral gradient that steepens from $-0.80$ to $-1.9$ in the west--east direction.

\subsection{Merger scenario}
\label{sec:mergerscenario}
The presence of an elongated radio halo connecting two radio relics in eFEDS~J093513.3+004746 and eFEDS~J093510.7+004910  indicates that the cluster is undergoing a major merger. This is supported by the galaxy density contour map in Figure \ref{fig_opt_0935.1} that shows two peaks in the north and south regions of the cluster system. In the case of a head-on binary collision on or close to the plane of the sky between sub-clusters that are equivalent in mass, radio relics tracing merger shocks are expected to be observed on the opposite sides of the cluster center \citep{VanWeeren2011}. However, RSE resides on the SE side of the merging axis that, as suggested by the major axes of the X-ray emission and the radio halo, is in the north--south direction. This can be explained by the off-axis merger of the cluster system where two sub-clusters eFEDS~J093513.3+004746 and eFEDS~J093510.7+004910 collide with a nonzero impact parameter \citep{VanWeeren2011}. Alternatively, the observed location of RSE that is not on the major axis of the halo can be explained in the shock re-acceleration scenario. In this scenario, a shock propagating towards the south re-accelerates a cloud of mildly relativistic electrons to generate the relic RSE which consequently would mark only the portion of the shock surface intercepting such a pre-existing cloud. The LOFAR and uGMRT observations do not detect the presence of extended radio emission in the western region of RSE which might be due to the sensitivity limit of these observations. Future deep X-ray and radio observations are needed to search for shock and radio emission at the locations of RSE and towards its western direction.


%
\section{Discussion and Conclusions}
\label{sec:sec4}
In this paper, we present the detection of a new supercluster that consists of eight clusters at a redshift of 0.36 observed in the eFEDS observations performed during the PV phase program. We use the X-ray data from eROSITA X-ray telescope, the optical data from the Hyper Suprime-Cam, and the radio data from the LOFAR and uGMRT to study the multi-wavelength properties of the cluster members of this supercluster and bridge regions in between them. 
This supercluster includes eight massive members where the northernmost clusters are going through an off-axis major merger activity. The optical and X-ray data suggest a triple merging system (double merger and a pre-merger)
with a mass ratio of 3:2:1. Further study of the radio emission in this region with LOFAR and uGMRT revealed the existence of two radio relics in the north and southeast regions of the clusters and an elongated radio halo indicative of an ongoing merger activity. In the filament direction, we observe a slight enhancement in the X-ray surface brightness and thermal electron number density  compared to the cluster emission along the off-filament axis, which might be an indication of an accretion along the putative filament and/or merger activity. However, the surface brightness profiles obtained from the eROSITA X-ray data are too shallow to further detect the density and temperature jumps at the location of relics, future deep follow-up X-ray observations are needed to reveal the details of the merger history of the major merger located in this supercluster.

We use eROSITA X-ray data to measure the integrated cluster properties, such as temperatures, luminosities, gas masses, and also total masses out to $R_{500}$  using a scaling relation. The X-ray properties of these clusters are similar to those of the common eFEDS cluster population (Ghirardini et al. in prep.) and lie along the $L_{X}-M$ and $M_{gas}-M$ scaling relations \citep{bulbul19}. Interestingly, the brightest member of the large-scale structure, which is also included in other optical and SZ catalogs, has a mass measurement that is consistent with the value reported in the \emph{Planck} catalog \citep{PSZ2}.  We further investigate the environmental influences of the large-scale structure on the evolution of the cluster members using the eROSITA data. By comparison to the dynamical and morphological properties of these clusters with the 338 confirmed clusters detected in the eFEDS field, we find that the morphological properties are consistent with the general eFEDS cluster population (Ghirardini et al. 2021). A larger statistical sample of the cluster population embedded in superclusters should be studied in order to characterize the environmental effects of the large-scale structure on the member morphology in detail, which will provide us with important clues as to the matter assembly and evolution of the large-scale structure in the Universe.

This work demonstrates the potential of SRG/eROSITA for not only mapping the large-scale structure of the Universe, but also probing the detailed physical properties of clusters embedded in the cosmic web that form superclusters. Assuming the Planck cosmology, our simulations (Comparat et al. in prep.) show that 
the number of superclusters with at least five members ---using the definition of supercluster in Sect.~\ref{sec:supercluster_definition}--- that is expected to be detected in the eROSITA\_DE area (the part of the sky that will be analyzed by the German eROSITA consortium, i.e., Galactic $l > 180 \textrm{~deg}$) at the final eROSITA all-sky survey (eRASS8) depth is 463, comprising a total of 3071 clusters. Comparing the properties of these clusters embedded in  superclusters with a general population of clusters of galaxies ($\sim 10^{5}$ of them are expected to be detected) observed in the X-ray, optical, and radio observations will allow us to statistically investigate the environmental  influence of the large-scale structure on the evolution of these clusters of galaxies.

\begin{acknowledgement}
The authors thank Joachim Trumper for useful discussions and comments on the manuscript. This work is based on data from eROSITA, the primary instrument aboard SRG, a joint Russian-German science mission supported by the Russian Space Agency (Roskosmos), in the interests of the Russian Academy of Sciences represented by its Space Research Institute (IKI), and the Deutsches Zentrum f{\"{u}}r Luft und Raumfahrt (DLR). The SRG spacecraft was built by Lavochkin Association (NPOL) and its subcontractors, and is operated by NPOL with support from the Max Planck Institute for Extraterrestrial Physics (MPE).

The development and construction of the eROSITA X-ray instrument was led by MPE, with contributions from the Dr. Karl Remeis Observatory Bamberg \& ECAP (FAU Erlangen-Nuernberg), the University of Hamburg Observatory, the Leibniz Institute for Astrophysics Potsdam (AIP), and the Institute for Astronomy and Astrophysics of the University of T{\"{u}}bingen, with the support of DLR and the Max Planck Society. The Argelander Institute for Astronomy of the University of Bonn and the Ludwig Maximilians Universit{\"{a}}t Munich also participated in the science preparation for eROSITA.

The eROSITA data shown here were processed using the eSASS software system developed by the German eROSITA consortium.

The Hyper Suprime-Cam (HSC) collaboration includes the astronomical communities of Japan and Taiwan, and Princeton University.  The HSC instrumentation and software were developed by the National Astronomical Observatory of Japan (NAOJ), the Kavli Institute for the Physics and Mathematics of the Universe (Kavli IPMU), the University of Tokyo, the High Energy Accelerator Research Organization (KEK), the Academia Sinica Institute for Astronomy and Astrophysics in Taiwan (ASIAA), and Princeton University.  Funding was contributed by the FIRST program from the Japanese Cabinet Office, the Ministry of Education, Culture, Sports, Science and Technology (MEXT), the Japan Society for the Promotion of Science (JSPS), Japan Science and Technology Agency  (JST), the Toray Science  Foundation, NAOJ, Kavli IPMU, KEK, ASIAA, and Princeton University. This paper makes use of software developed for the Large Synoptic Survey Telescope. We thank the LSST Project for making their code available as free software at  http://dm.lsst.org This paper is based in part on data collected at the Subaru Telescope and retrieved from the HSC data archive system, which is operated by Subaru Telescope and Astronomy Data Center (ADC) at NAOJ. Data analysis was in part carried out with the cooperation of Center for Computational Astrophysics (CfCA), NAOJ.

The Pan-STARRS1 Surveys (PS1) and the PS1 public science archive have been made possible through contributions by the Institute for Astronomy, the University of Hawaii, the Pan-STARRS Project Office, the Max Planck Society and its participating institutes, the Max Planck Institute for Astronomy, Heidelberg, and the Max Planck Institute for Extraterrestrial Physics, Garching, The Johns Hopkins University, Durham University, the University of Edinburgh, the Queen’s University Belfast, the Harvard-Smithsonian Center for Astrophysics, the Las Cumbres Observatory Global Telescope Network Incorporated, the National Central University of Taiwan, the Space Telescope Science Institute, the National Aeronautics and Space Administration under grant No. NNX08AR22G issued through the Planetary Science Division of the NASA Science Mission Directorate, the National Science Foundation grant No. AST-1238877, the University of Maryland, Eotvos Lorand University (ELTE), the Los Alamos National Laboratory, and the Gordon and Betty Moore Foundation.

LOFAR data products were provided by the LOFAR Surveys Key Science project
(LSKSP; https://lofar-surveys.org/) and were derived from observations with the International LOFAR Telescope (ILT). LOFAR (van Haarlem et al. 2013) is the Low Frequency Array designed and constructed by ASTRON. It has observing, data processing, and data storage facilities in several countries, that are owned by various parties (each with their own funding sources), and that are collectively operated by the ILT foundation under a joint scientific
policy. The efforts of the LSKSP have benefited from funding from the European Research Council, NOVA, NWO, CNRS-INSU, the SURF Co-operative, the UK Science and Technology Funding Council and the J\"{u}lich Supercomputing Centre.

We thank the staff of the GMRT that made these observations possible. GMRT is run by the National Centre for Radio Astrophysics of the Tata Institute of Fundamental Research.

VB acknowledges support by the DFG project nr. 415510302. AB and DNH {acknowledge support from the ERC through the grant ERC-Stg DRANOEL n. 714245}. AB acknowledges support from the MIUR grant FARE ``SMS''. MB acknowledges support from the Deutsche Forschungsgemeinschaft under Germany's Excellence Strategy - EXC 2121 "Quantum Universe" - 390833306. 
This work was supported in part by World Premier International Research Center Initiative (WPI Initiative), MEXT, Japan, and JSPS KAKENHI Grant Number JP19KK0076.
{A.Botteon} acknowledges support from the VIDI research programme with project number 639.042.729, which is financed by the Netherlands Organisation for Scientific Research (NWO). 
RJvW acknowledges support from the ERC Starting Grant ClusterWeb 804208. GB and RC acknowledge support from INAF through mainstream program "Galaxy cluster science with LOFAR".

\end{acknowledgement}

\bibliographystyle{aa} 
\bibliography{references} 

\begin{appendix}
\section{Morphological parameters distribution}
\begin{figure}[!h]
    \centering
    \includegraphics[width=0.5\textwidth]{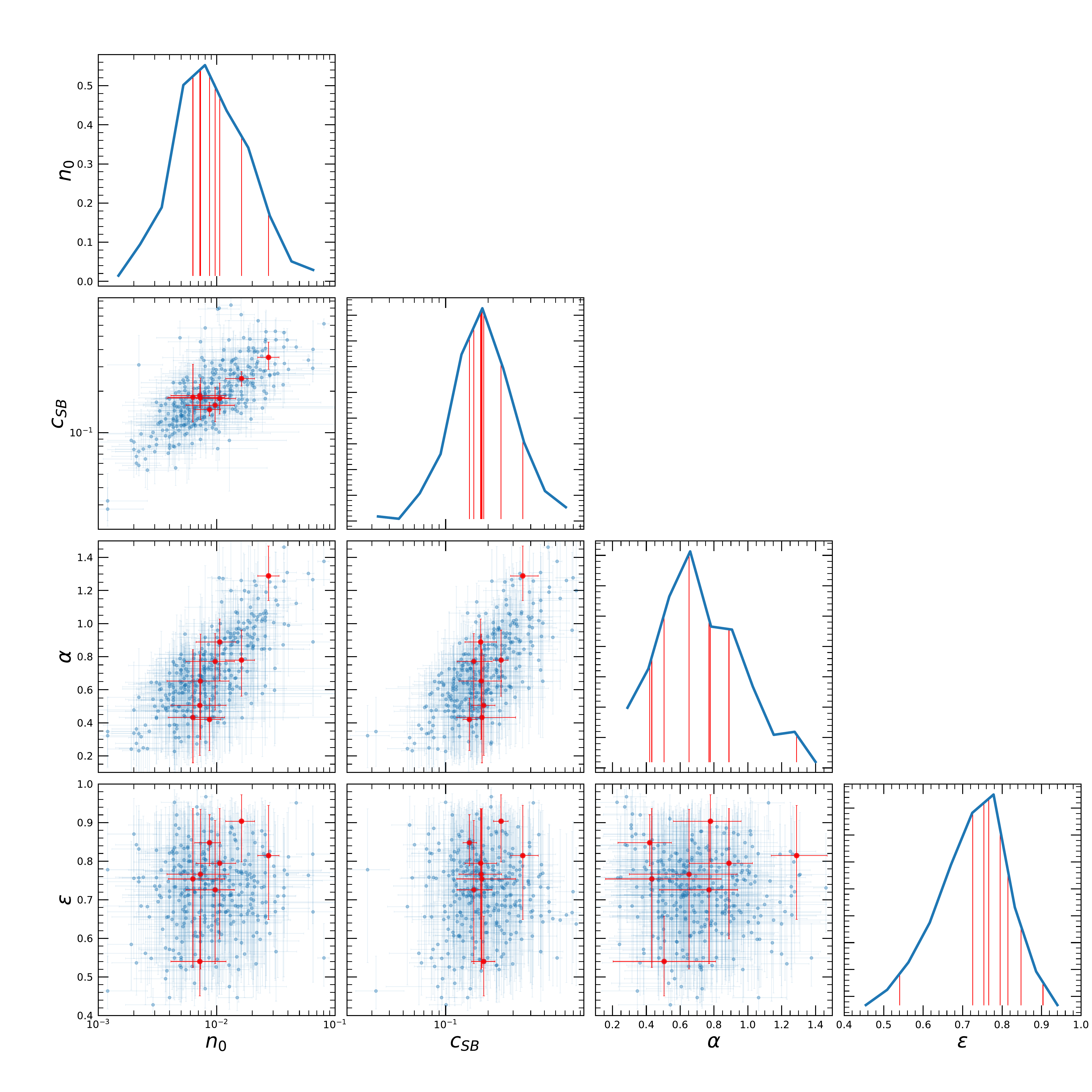}
    \caption{Morphological parameter distribution for the clusters in eFEDS field. In blue we show the location in the distribution for the clusters in the entire eFEDS field, and in red we highlight the distribution for the cluster members of the supercluster.}
    \label{fig_morph}
\end{figure}

\begin{figure*}
\includegraphics[width=0.5\textwidth]{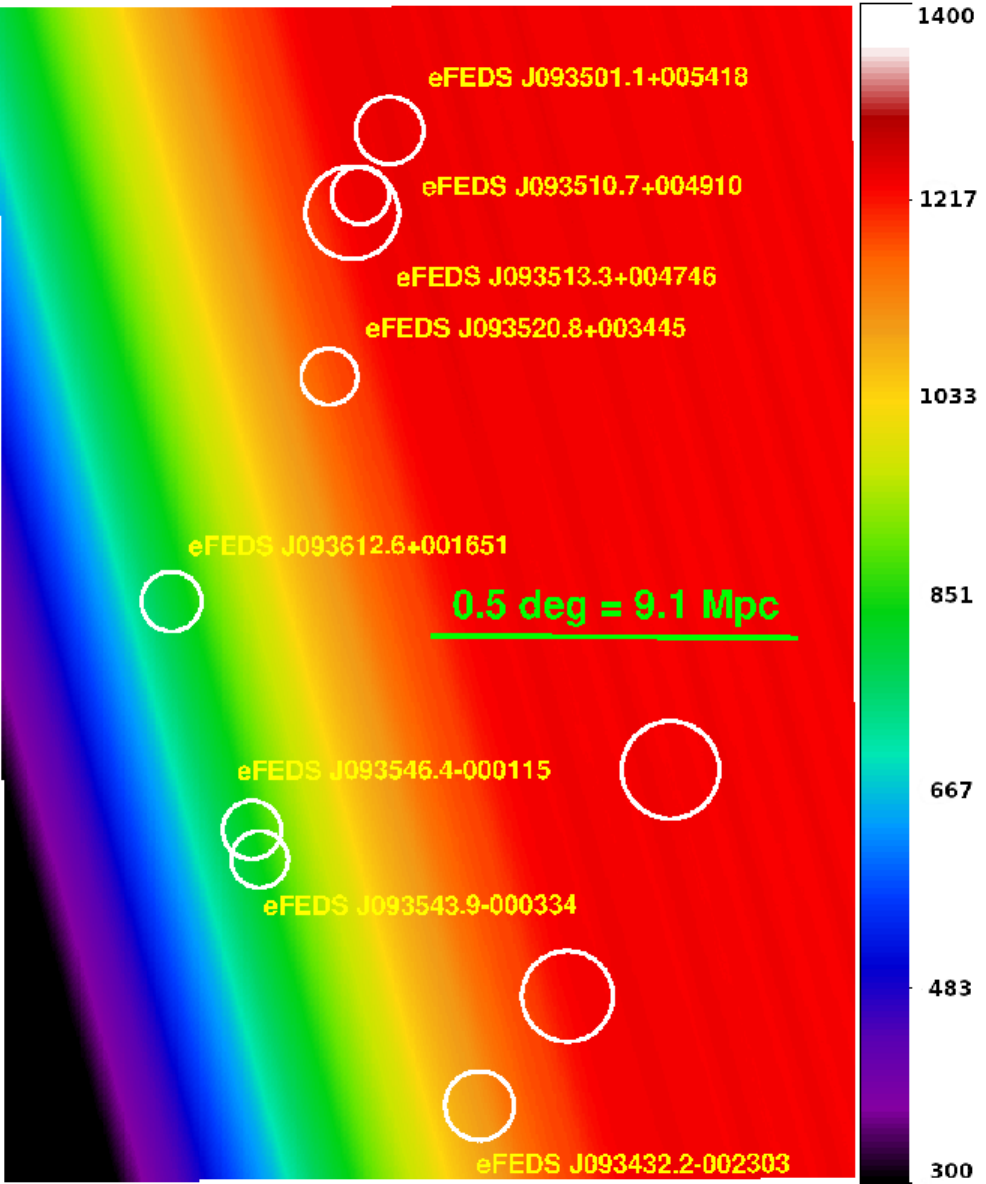}~
\includegraphics[width=0.5\textwidth]{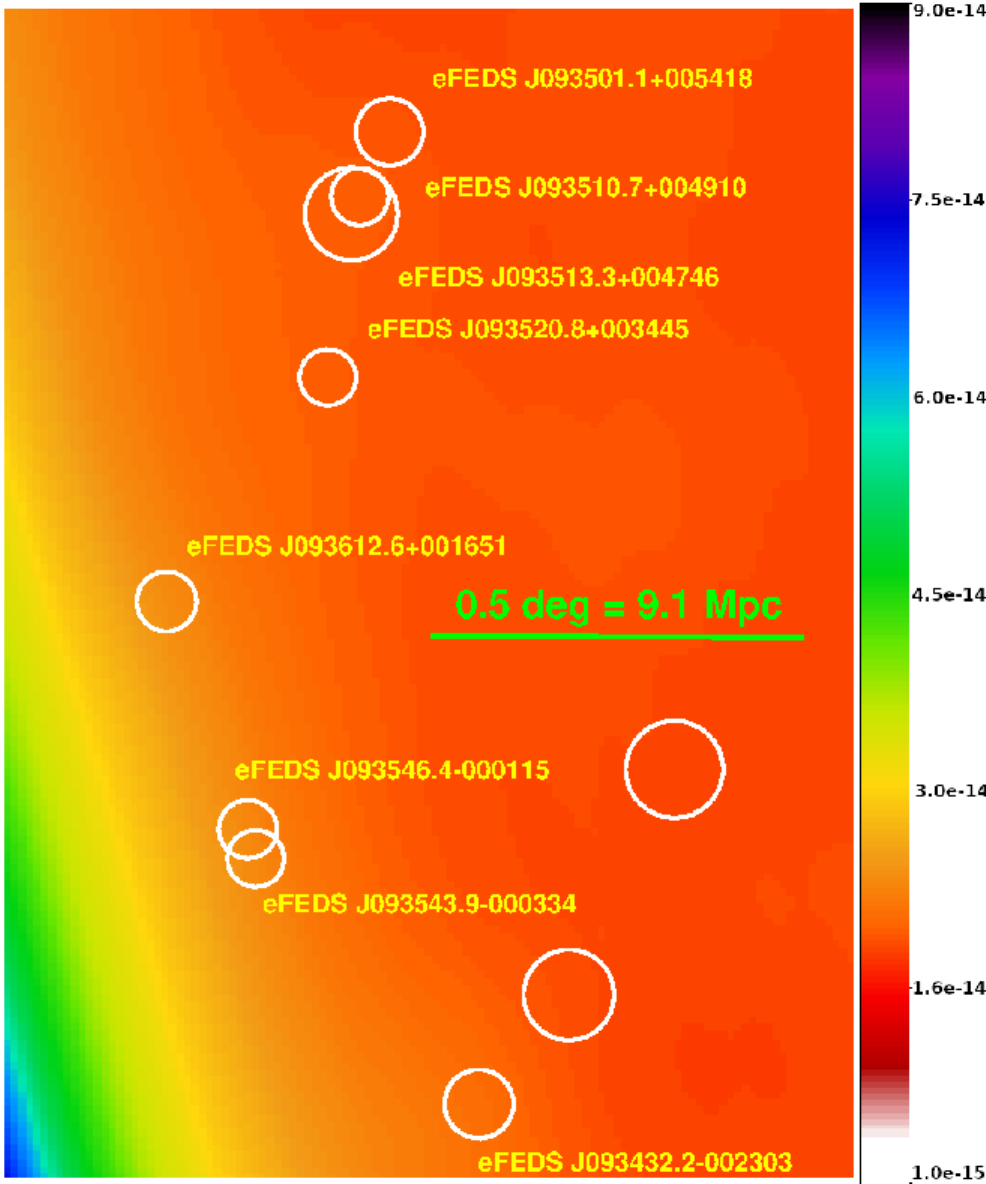}~
\caption{ \emph{Left}: eROSITA exposure map in the soft [0.5-2.0] keV energy band. \emph{Right}: eROSITA sensitivity map in units of $\textrm{erg}\textrm{s}^{-1} \textrm{cm}^{-2}$.}
\label{fig_sens_exp}
\end{figure*}

\end{appendix}

\end{document}